\documentclass[twocolumn,10pt]{article}
\usepackage[utf8]{inputenc}
\usepackage{graphicx,amssymb,amsmath}
\usepackage{pdfpages}

\def\fig#1{Fig.~\ref{#1}}
\def\be{\begin{equation}}
\def\ee{\end{equation}}
\def\beq{\begin{equation}}
\def\eeq{\end{equation}}

\definecolor{mygray}{rgb}{0.45, 0.45, 0.6}

\usepackage[a4paper,
            bindingoffset=0.2in,
            left=0.6in,
            right=0.6in,
            top=0.6in,
            bottom=0.8in,
            footskip=.25in]{geometry}

\title{Protein overexpression can induce the elongation of cell membrane nanodomains}

\author{Julie Cornet$^{a}$, Pascal Preira$^{b}$, Laurence Salom\'e$^{b}$, Fr\'ed\'eric Daumas$^{b}$,\\ 
	Bernard Lagane$^{c}$, Nicolas Destainville$^{a}$, Manoel Manghi$^{a}$$^{*}$, Fabrice Dumas$^{b}$$\dag$\\
        \small $^{a}$Laboratoire de Physique Th\'eorique,  Universit\'e Toulouse III - Paul Sabatier, CNRS, Toulouse, France \\
        \small $^{b}$Institut de Pharmacologie et de Biologie Structurale (IPBS), Universit\'e de Toulouse,\\ 
        \small CNRS, Universit\'e Toulouse III - Paul Sabatier (UPS), Toulouse, France \\
        \small $^{c}$Infinity, Universit\' e de Toulouse III - Paul Sabatier, CNRS, Inserm, Toulouse, France \\\\
        \small $^{*}$\tt{manoel.manghi@univ-tlse3.fr} \qquad $^\dag$\tt{fabrice.dumas@univ-tlse3.fr} \\
}

\begin{document}
\maketitle
\begin{abstract}             
In cell membranes, proteins and lipids are organized into sub-micrometric nanodomains of varying size, shape and composition, performing specific functions. Despite their biological importance, the detailed morphology of these nanodomains remains unknown.  Not only can they hardly be observed by conventional microscopy due to their small size, but there is no full consensus on the theoretical models to describe their structuring and their shapes. Here, we use a combination of  analytical calculations and Monte Carlo simulations based upon a model coupling membrane composition and shape to show that increasing protein concentration leads to an elongation of membrane nanodomains. The results are corroborated by Single Particle Tracking measurements on HIV receptors, whose level of expression in the membrane of specifically designed living cells can be tuned. These findings highlight that protein abundance can modulate nanodomain shape and potentially their biological function. Beyond biomembranes, this meso-patterning mechanism is of relevance in several soft-matter systems because it relies on generic physical arguments. 
\end{abstract}

\section*{Introduction}

The plasma membrane forms a hydrophobic barrier to separate the interior from the exterior of cells and a two-dimensional fluid matrix for proteins. 
Membrane components self-organize spontaneously to maximize molecular interactions in order to reach the lowest free energy state. The first attempt to describe membrane organization by Singer and Nicholson is now 50 years old~\cite{Singer1972}, and many updates have been proposed since then~\cite{Jacobson1995,Vereb2003,Kusumi2012,Garcia-Parajo2014,Bernardino2016,Sezgin2017}. 
Lipid--lipid, protein--protein and lipid--protein interactions lead to the formation of domains in which these constituents are unevenly distributed. This structuring is complicated by the interaction with the protein scaffold of the inner membrane surface and the involvement of membrane associated elements~\cite{Goyette2017}. Among the huge diversity of membrane domains, the most studied ones are certainly lipid rafts described as heterogeneous, dynamic, and short-lived cholesterol- and sphingolipid-enriched membrane nanodomains (10-200~nm) that are in a liquid ordered phase~\cite{Simons1997}. Beyond rafts, it is now widely understood that nanoscale clustering is a common feature of membrane proteins. This results in a cell membrane with composition and physical properties that are different from average membrane properties. A number of studies have reported the involvement of nano-domains in biological functions such as signal transduction~\cite{Simons2000}, regulation of membrane trafficking~\cite{DiazRoher2014}, immune signalling~\cite{Sezgin2017,Goyette2017}, or infectious processes~\cite{Dumas2017,Waheed2009,Yang2015,Yang2017,Peruzzu2022}.

Over the past 20 years,  super-resolution microscopy techniques among which Single Particle Tracking (SPT) have revealed that these nanodomains have dimensions ranging from a few to several hundred nanometers~\cite{Jacobson2019,Kusumi2011}.
SPT allows to follow the dynamics of individual molecules in living cells with unique accuracy over dozens of seconds [for review see~\cite{Alcor2009,Manzo2015,Jacobson2019}]. While the behavior of a single trajectory might be stochastic, the statistical analysis of many trajectories provides insight on properties such as membrane receptor activation, assembly/dissociation of signalling clusters, or protein reorganization due to virus interactions~\cite{Daumas2003,Ruthardt2011,Tsunoyama2018,Mascalchi2012,Lee2019,Baker2007_1}. 
\begin{figure}[t!]
\centering
\includegraphics[width=\columnwidth]{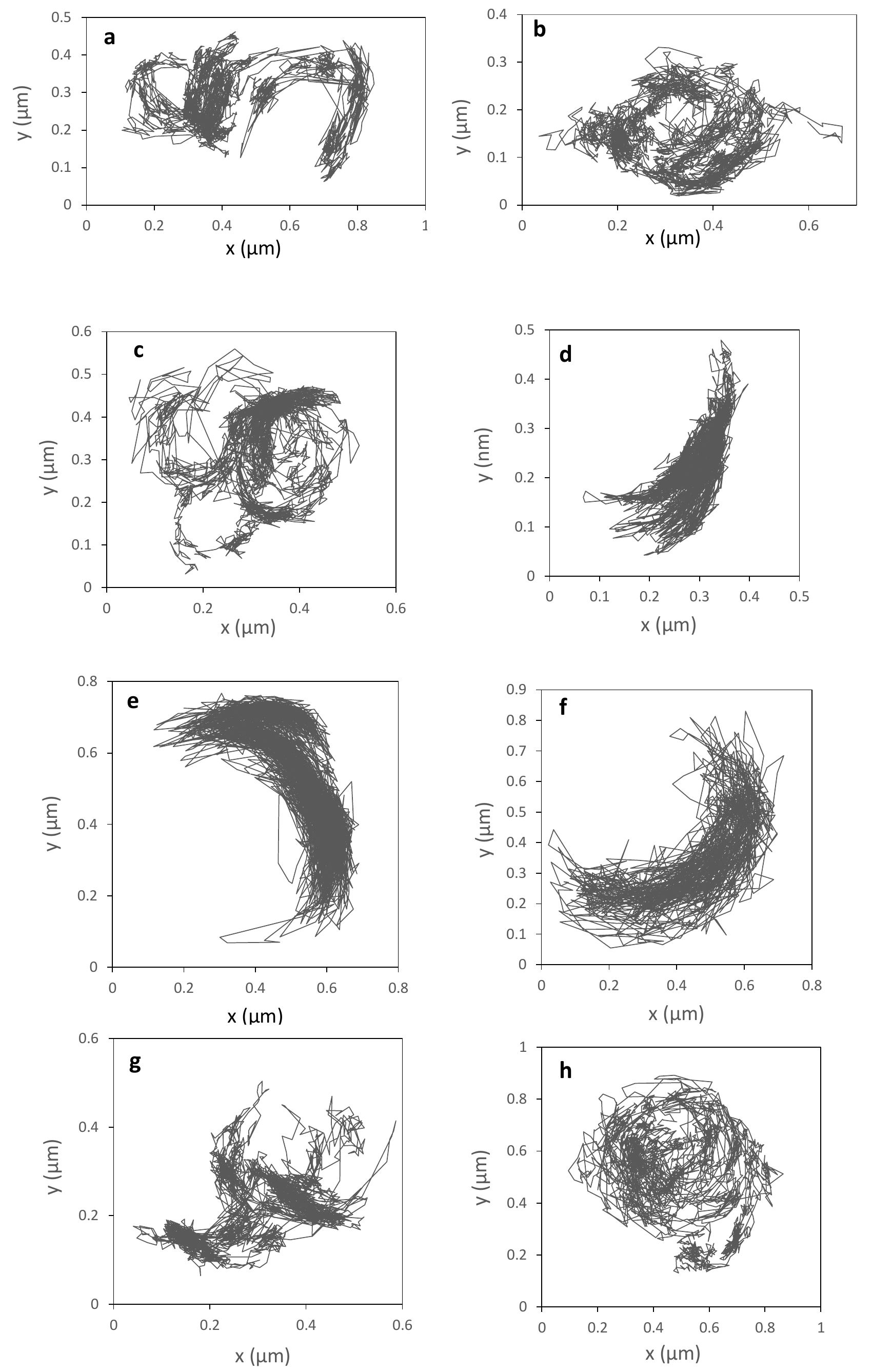}
\caption{Example of oddly shaped MOR trajectories acquired by SPT on NRK fibroblasts. Trajectories have been acquired in the absence of ligand (a-b) or in the presence of 1~$\mu$M of agonist ligand DAMGO (c-h). Out of 100 trajectories without ligand, 2 are curled (2~\%) whereas in presence of DAMGO it increases to 9 over 60 (15~\%).}
\label{fig:MOR}
\end{figure}
SPT permits to classify trajectories in different categories than can be correlated to various physiological events. For instance, proteins that have no specific interaction with any membrane structure present a Brownian motion, those interacting with cytoskeleton filaments display a directional motion and those that have an affinity for micro-domains present a permanently or transiently confined diffusion~\cite{Baker2007}. For many years, in different studies, we have observed some oddly shaped confined SPT trajectories that we could not correlate to physical or physiological events. We give an illustrative example of these atypical trajectories that mostly presented an elongated or curved horseshoe shape in \fig{fig:MOR}. The measurements were obtained by tracking the Mu Opioid Receptor (MOR) at basal state or in the presence of an agonist named DAMGO. Interestingly, we have observed that the proportion of these elongated and/or curled trajectories seemed to increase in the presence of the agonist as shown in \fig{fig:MOR}. It is known that the binding of ligands induces conformational changes of GPCRs~\cite{Salamon2000} and can alter the oligomeric arrangement of receptors~\cite{Xuet2019}. More recently, Civciristov and co-authors stated that ``DAMGO or morphine induces the assembly of different protein-interaction networks''~\cite{Civciristov2019}. This might induce a local increase of protein concentration and reshape membrane domains, in agreement with the observation that some nanodomains show more elongated contours when membrane proteins are overexpressed~\cite{private_com}. In the same idea, Merklingler \textit{et al.} revealed by super-resolution microscopy that increasing the expression level of syntaxin induces an increase of the local protein concentration and elongation of the nanodomains in which it is confined [compare their Fig.~5 and 5S in~\cite{Merklinger2017}]. Taken together, these observations led us to hypothesise that the increased local concentration of membrane proteins can induce a modification of the shape of the domains that enclose them.

To verify this hypothesis, we adopt in parallel theoretical and experimental approaches.
\begin{figure}[t!]
\centering
\includegraphics[width=\columnwidth]{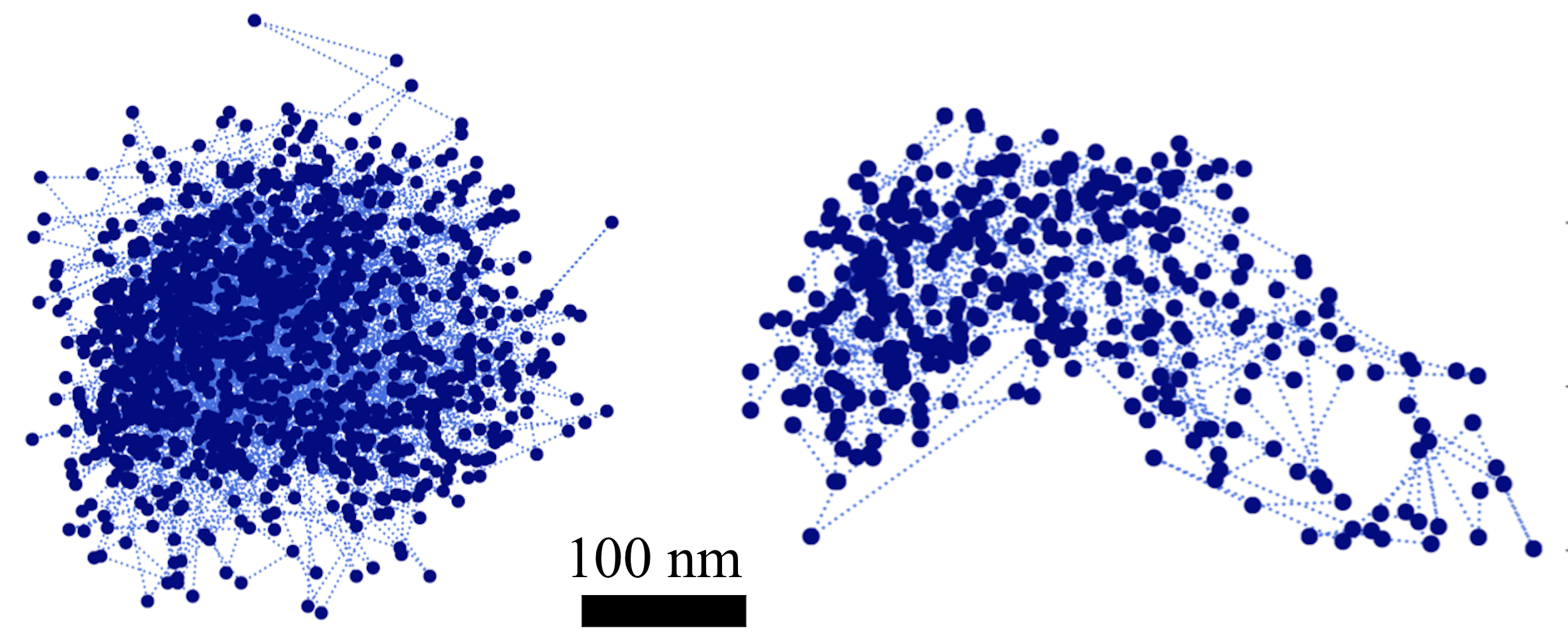}
\caption{Examples of SPT trajectories of CCR5 receptors confined into nanodomains at the surface of Affinofile cells (see text). More roundish nanodomains are observed when the proteins have a low expression level (left) and more elongated ones when the proteins are overexpressed (right).}
\label{fig:traj}
\end{figure}
In our theoretical framework, membrane meso-patterning ensues from the competition between (i) short-range attractive forces promoting a condensed phase of proteins and (ii) an additional mechanism, related to the coupling between the membrane composition and its shape (curvature), making too large domains unstable in the favor of smaller ones in equilibrium~\cite{Komura2014,Schmidt2017,review}. We performed mesoscale numerical simulations~\cite{Cornet2020}, where a vesicle made of two species having different spontaneous curvatures is simulated with a Monte Carlo (MC) algorithm coupling the composition and the membrane elasticity, see Supplemental Material (SM). Below the critical temperature, it leads to the formation of nanodomains, instead of a macrophase separation. As a function of the parameter values, we observe the formation of more or less elongated nanodomains. These simulations are guided by analytical calculations indicating the range of physical parameters entering the numerical model for which the elongation mechanism occurs. 
In these calculations, we use a simplified model of pairwise mid-range repulsive forces because it has been demonstrated that the coupling between composition and curvature plays the same role as such a pairwise repulsive force, from a statistical physics point of view~\cite{Weitz2013}: both mechanisms make large domains unstable because their energy grows faster than their size. Since we have also an expertise in the study of the dynamics of HIV receptors, we have in parallel performed SPT experiments (\fig{fig:traj}) on three transmembrane proteins involved in HIV-1 infection: CD4, CCR5 and CXCR4~\cite{Lodowski2009,Dumas2017}. It has been shown that their distribution at the surface of  target cells is heterogeneous~\cite{Viard2002,Heredia2007,Mulampaka2011,Singer2001,Jung2016}, and proposed that membrane nanodomains, by concentrating the receptors, would be the preferential sites of virus entry into cells~\cite{Steffens2003,Gaibelet2006,Dumas2017}.  We have used the 293-Affinofile cell line to record receptor trajectories at different expression levels. These cells, in which CD4 and CCR5 expression can be independently induced~\cite{Johnston2009}, were further transduced to stably express low or high level of CXCR4. We have recorded the trajectories of these proteins at different expression levels. Our observations reveal that an elongation of membrane nanodomains occurs upon protein overexpression consistently with our theoretical findings.

\section*{Materials and Methods}

\subsection*{Cell culture and transfection}
Affinofile cells, provided by Dr. B. Lee, Mont Sina\"i hospital, New York, NY, USA, were maintained in Dulbecco's modified Eagle's medium with 10\% dialysed fetal calf serum supplemented with 50$\mu$g/ml blasticidin (D10F/B). 24-well plates were seeded with $1.2\times10^5$ Affinofile cells/well, and expression of CD4 and CCR5 was induced the following day at 37$^\circ$C with various concentrations of minocycline and/or ponasterone, respectively, for 18h~\cite{Johnston2009}.

Stable expression of CXCR4 was performed through lentiviral transduction of 293-Affinofile cells with the pTRIP deltaU3 lentiviral vector encoding the CXCR4 sequence. Lentiviral particles were prepared by transient cotransfection of HEK 293T cells with the pTRIP lentiviral vector [a gift from P. Charneau, Institut Pasteur, Paris, originally described in~\cite{Zennou2001}] encoding the CXCR4 sequence, the p8.71 encapsidation plasmid and a plasmid encoding the VSV envelope glycoprotein G (pVSVG) (at a 2/2/1 ratio). Transfection was performed using a standard calcium phosphate precipitation method. Forty-eight hours later, lentiviral particles were harvested and quantified using a HIV capsid protein ELISA kit. Then, 293-Affinofile cells were inoculated by different amounts of lentiviral particles so as to obtain cells with different transduction efficiencies. From the different sets of transduced cell populations, cell clones expressing low or high levels of CXCR4 were further selected using serial dilutions in 96-well culture plates. During all these experiments, CXCR4 expression was monitored by flow cytometry using the anti-CXCR4 mAb 12G5 labeled with phycoerythrin.

\subsection*{CD4, CCR5 and CXCR4 quantification}
The number of CD4, CCR5 and CXCR4 proteins per cell, as displayed in Table~\ref{Table1}, have been estimated by quantitative fluorescence-activated cytometry (qFACS) using Quantibrite PE calibration beads (BD quantibrite beads cat. number 340495). BD Quantibrite PE is a set of beads covalently conjugated with four levels of phycoerythrin (PE). By running a BD Quantibrite PE tube at the same instrument settings as the assay, the fluorescence intensity can be converted into the number of PE molecules (i.e. antibody) bound per cell~\cite{Davis}. 100 $\mu$L of the Affinofile cells in suspension (1$\times10^5$) have been incubated with 3 $\mu$L of phycoerythrin-conjugated anti-CD4, anti-CCR5 or anti-CXR4 for 60~min at 4$^\circ$C in the dark and in the presence of the ATPase inhibitor sodium azide (0.1\%) to avoid internalization. Cells were then washed twice with PBS, resuspended in 100 $\mu$L of PBS and immediately analyzed by flow cytometry without fixation. All experiments have been carried out in triplicate.
\begin{table}[t]
\centering
\begin{tabular}{lcc}
number of proteins  & low & high  \\
\hline
 CD4  & $2800 \pm 300$ & $100000\pm 13000$\\
CCR5 & $3600\pm 800$ & $125000 \pm 8000$\\
CXCR4 &	$15 000 \pm 2000$ & $120 000 \pm 7000$\\
\hline
\end{tabular}
\caption{\label{Table1} Average number of proteins per cell in the low and high expression levels for the three types of HIV receptors. Low expression of CD4 and CCR5 conditions correspond to cells that were not induced by minocycline and/or ponasterone. High CD4 and CCR5 expression conditions correspond to those induced with 5~ng/ml minocycline and/or 1~$\mu$M ponasterone, respectively.}
\end{table}

\subsection*{Single Particle Tracking experiments}
To be able to acquire long time trajectories we have used antibodies that present higher affinities than Fab fragments. Since CD4, CCR5 and CXCR4 were known to interact with each other, we have used antibody clones that had been validated in the literature and described as not to interfere in those interactions (when the information was available): T21/8 for CCR5~\cite{Mascalchi_these2012}, OKT4 for CD4~\cite{Staudinger2003} and 12G5 for CXCR4.
Cells were plated on coverslips previously incubated with 0.1~mg/ml poly-L-lysine for 5~min. For a homogenous attachment of the cells, a gentle centrifugation (50~g, 7~min) was performed. Proteins (CD4, CCR5 or CXCR4) were labeled for 15~min with 0.03~nM biotinylated antibodies coupled to 0.3~nM fluorescent (655~nm) streptavidin-coated Quantum Dots (QD) (Molecular Probes, LifeTechnologies). This antibody/QD ratio of 1:10 ensures that no more than one antibody is bound per particle to prevent that one particle interacts with multiple receptors on the cell surface~\cite{Mascalchi2012_2}.
Tracking and observations were performed at room temperature on an Axioplan 2 microscope (Zeiss) equipped with a Cascade II 512 EM-CCD camera (Roper Scientific) operating at  a 25~Hz acquisition frequency. The fluorescent nanoparticles were illuminated with an X CITE 120 light source containing a metal halide vapor short arc lamp (Exfo), and observed through a fluar 100$\times$/1.30 oil UV objective associated to a 1.6$\times$ multiplier tube lens in front of the camera. Acquired video durations go from 30~s to 80~s. The maximum duration of measurements of a slide has been fixed to 30~min in order to avoid cellular stress. In these conditions, 1 to maximum 4 trajectories could be acquired on a same cell.

As any experimental measurement, SPT suffers from a small, but finite localization error. To experimentally bound the average error made during the position determination, we  immobilized quantum dots on a glass coverslip and embedded them in a 15\% polyacrylamide gel, recorded 80~s videos and determined the positions of each particle in each image. The pointing error was defined in $x$ and $y$ as the standard deviation of the position distribution of the immobilized quantum dots. We obtained a value of 7 nm in both $x$ and $y$ coordinates (Fig.~S5). The thermal drift of the microscope is \textit{de facto} taken into account by this measurement.

\subsection*{Single Particle Tracking trajectory analysis}
The trajectories analysis is based on the calculation of the Mean Square Displacement (MSD)~:
\begin{equation}
{\rm MSD}(s)=\langle [ \mathbf{r}(t+s)-\mathbf{r}(t)]^2 \rangle 
\end{equation}
where $\mathbf{r}$ is the 2D tracked particle position and the average is taken over the successive frames at successive times $t$. It characterizes the displacement during frames separated by a duration $s$. 
MSD plots are a useful tool, having been used for several decades to characterize the diffusional behavior of tracked objects in SPT experiments ~\cite{Daumas2003,Quian1991,Mascalchi2012,Meilhac2006,Michalet2010}. They allow us to classify the diffusive behaviors of trajectories or parts of trajectories into different categories namely random, confined, walking confined or directed diffusion~\cite{Baker2007}, according to the mathematical expression of the MSD. To do so, we systematically determined, by using a non-linear least mean-square regression algorithm based on a $\chi^2$ test, which equation gave the best fit of the MSD versus time plots~\cite{Daumas2003} between:
\begin{itemize}
	\item free diffusion in 2 dimensions, with diffusion coefficient $D$
\be
{\rm MSD}(s)=4Ds
\ee 
	\item diffusion confined in a domain 
\be
{\rm MSD}(s) \simeq 2 \Delta \mathbf{r}^2 [1-\exp(-s/\tau)]
\ee 
where the relaxation time $\tau = \Delta \mathbf{r}^2/(2D)$; 
	\item directed diffusion
\be
{\rm MSD}(s)=4Ds + v^2 s^2
\ee
where $v$ is the drift or transport velocity;
	\item walking confined diffusion
\be
{\rm MSD}(s)=A\left[1-\exp\left(-4D_{\rm micro}s/A\right)\right]+4D_{\rm macro}s
\ee 
where $A$ is the characteristic area of the confined region, $D_{\rm macro}$ is the long time diffusion and $D_{\rm micro}$ is the diffusion coefficient inside the domain.
\end{itemize}
It has to be noticed that no walking confined diffusion trajectories (diffusion in a confined domain, itself diffusing) have been observed here.
For more details, \fig{fig:traj} provides examples extracted from the SPT trajectories studied in this work.

\subsection*{Confinement index}
\label{conf:index}
\begin{figure}[t!]
\centering
\includegraphics[width=0.9\columnwidth]{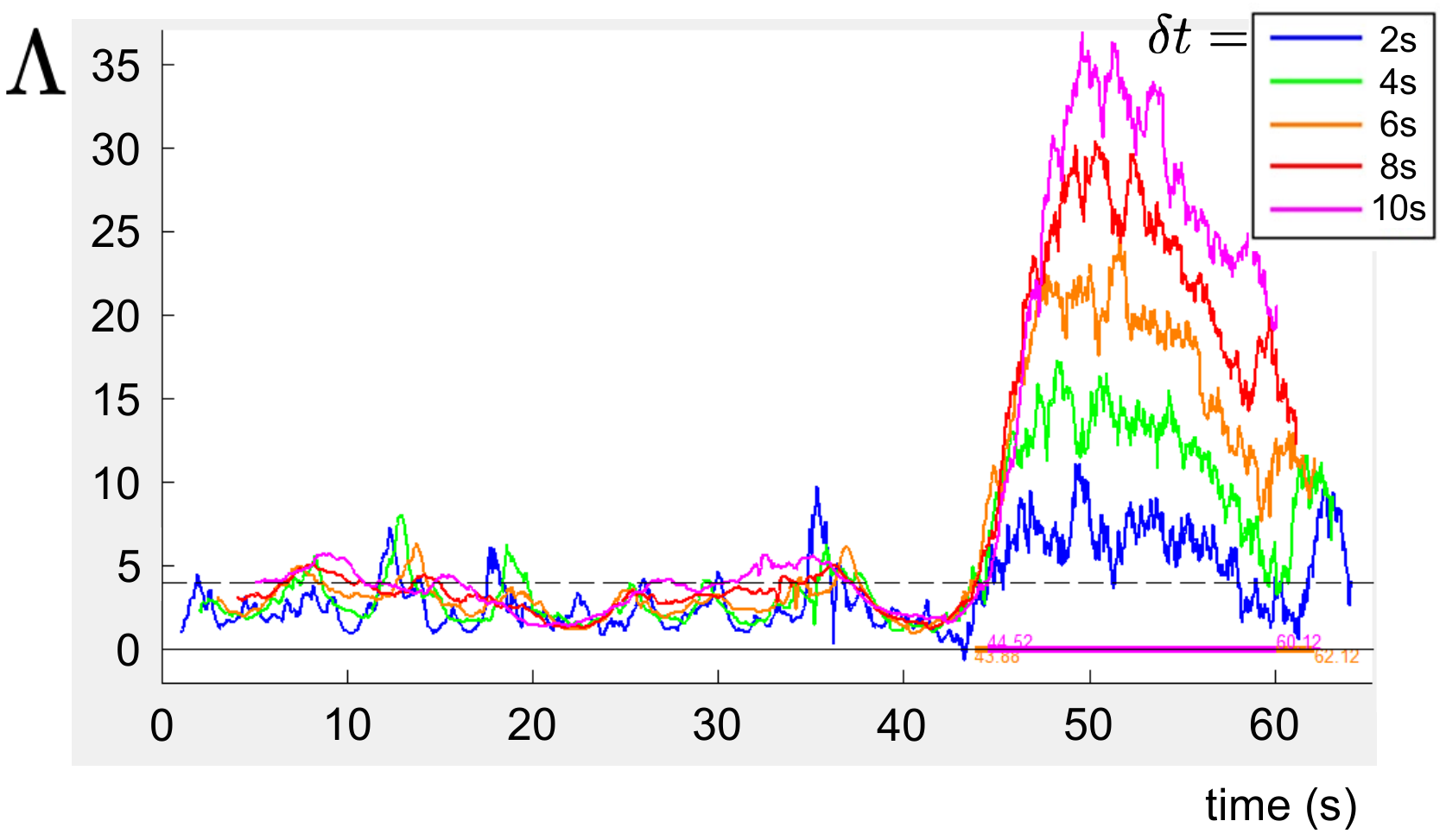}
\includegraphics[width=0.9\columnwidth]{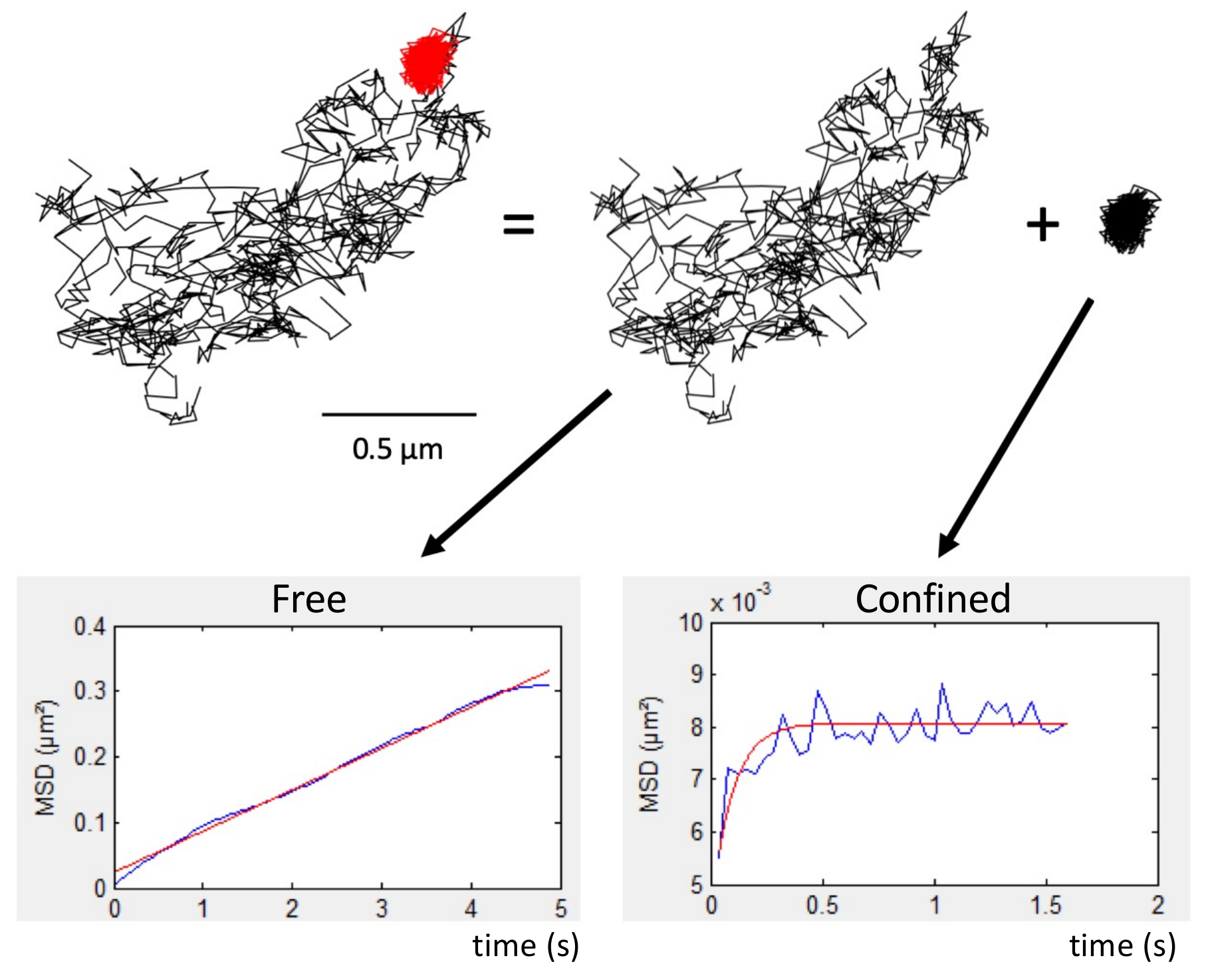} 
\caption{Top: Example of use of the confinement index $\Lambda$ on one of our SPT experimental trajectories, in function of time. It is calculated over a sliding time window of duration $\delta t$ given by the color code. By the end of the trajectory, the index becomes larger than the threshold 4 (dashed line), indicating a marked confinement zone~\cite{Meilhac2006}. The colored line below the plots represents the duration of the confinement. Middle: the so-obtained transient confinement zone is represented in red on the SPT trajectory which is split into two parts, presumably a free random walk and the confined transient confinement zone. Bottom: This is confirmed by inspection of the MSD plots, the first one is linear and the second one is typical of diffusion confined in a nanodomain. The time is in s on all axes.}
\label{fig:MSD}
\end{figure}

In Ref.~\cite{Meilhac2006}, some of us have developed and characterized a tool to detect transient confinement in single-molecule membrane trajectories. Basically, it consists in identifying, along a trajectory, sets of $n\gg1$ successive positions $\mathbf{r}_i$ [of duration $\delta t(n)$], the variance $\Delta \mathbf{r}^2(n)$ of which is significantly smaller than what it would be if the diffusion were free, $\Delta \mathbf{r}^2(n) \ll 4 D \delta t(n)$. The index $\Lambda=D \delta t(n)/\Delta \mathbf{r}^2(n)$ must be large enough (larger than 4 in practice), for a long enough duration, to rule out the possibility of statistical fluctuations of a non-confined trajectory and to ascertain confined diffusion. The diffusion coefficient $D$, that can depend on the membrane region where the tracked molecule diffuses, must be monitored in real time with the help of a local MSD plot. This tool has been shown to be robust and to detect transient confinement zones with a good accuracy.  A comprehensive example is given in \fig{fig:MSD}. Transient confinement zones that are isolated through this procedure are considered as independent measurements.

\subsection*{Reagents and antibodies}
Minocycline (Sigma Aldrich, St. Louis, MO, USA) was dissolved in dimethyl sulfoxide to generate a stock concentration of 1~mg/mL. Ponasterone (Invitrogen, Carlsbad, CA, USA) was dissolved in 100\% ethanol to generate a stock of 1~mM. Blasticidin HCl (Invitrogen, Carlsbad, CA, USA) was dissolved in sterile water to generate a stock solution of 5~mg/mL.
For protein quantification the following phycoerythrin-conjugated IgGs antibodies have been used: anti-CD4 (OKT4,  Biolegend), anti-CCR5 (2D7, BD Biosciences, San Jose, CA, USA), anti-CXCR4 (12G5,  Biolegend).
For SPT experiments the following biotinylated antibodies have been used: anti-CD4 (OKT4, Biolegend, Ref. 317406), anti-CCR5 (T21/8, Biolegend), anti-CXCR4 (12G5, Biolegend).

\section*{Results}

\subsection*{Elongated domains are more stable than circular ones above a certain size}
We first study analytically the shape of one membrane nanodomain in thermodynamical equilibrium.
Its stability comes from the action of two competitive interactions. A short-range attraction between components of the same nature and/or repulsion between components of different nature~\cite{Schmidt2008,Bories2018} gives rise to a line tension at the boundary between two phases. A mid-range repulsive interaction due to the coupling between membrane composition and membrane elasticity prevents macrophase separation~\cite{Stradner2004}. We study for which range of the physical parameters an elongated shape is more stable than a circular one. The model considers generically an elliptic nanodomain of semi-axes $r_0 a$ and $r_0/a$, with $a\geq1$, so that the ellipse has area $A=\pi r_0^2$. Its aspect ratio AR, defined as the ratio of major to minor axes, is ${\rm AR} = a^2$. 
The total energy of the system is $E_{\rm tot}(a) =  E_{\rm bulk}+ E_{\rm rep} + E_{\rm line}$ where $E_{\rm bulk}\propto r_0^2$ is the cohesive energy of the nanodomain due to inter-molecular short-range forces, and is simply proportional to its area. $E_{\rm bulk}$ does not depend on $a$ and will be skipped in the following calculations. 
The repulsion energy between the membrane components inside the nanodomain is supposed to be pairwise and to have a finite range $\xi$:
\begin{equation}
E_{\rm rep} =E_0 \int_{A\times A}\rho(\mathbf{r})\,\varphi \left( \frac{|\mathbf{r} -\mathbf{r}'|}{\xi}\right)\rho(\mathbf{r}') {\rm d}\mathbf{r} \, {\rm d}\mathbf{r}'
\end{equation}
where $\varphi$ is the interaction potential, and $\rho(\mathbf{r})$ the particle density inside the nanodomain. The parameter $E_0$ sets the strength of the repulsion. In the case where this repulsive energy comes from the coupling with the membrane curvature, the molecules in the nanodomain induce a spontaneous curvature $C_0$ different from the average one (assumed to be 0 for simplicity sake). It has been shown that the screening length is $\xi = \sqrt{\kappa/\sigma}$ and that $E_0 \propto \sigma C_0^2$ in the low tension limit~\cite{Weitz2013,Cornet2020}. Here $\kappa$ and $\sigma$ are respectively the membrane bending modulus and surface tension. In principle, the function $\varphi$ decays exponentially at long distances, being for example a Bessel function. In order to get an analytically tractable model, we assume it to be the Gaussian $\varphi({|\mathbf{r} -\mathbf{r}'|}/{\xi})=\exp[-(\mathbf{r} -\mathbf{r}')^2/(2\xi^2)]$, which is sufficient at the scaling level. 
We also suppose that $\rho(\mathbf{r})=\frac12 \exp\left[-\frac12\left(\frac{x^2}{(r_0a)^2} + \frac{(a y)^2}{r_0^2}\right)\right]$ is a Gaussian density with $\mathbf{r}=(x,y)$. The prefactor $1/2$ ensures 
$\int_{\mathbb{R}^2} \rho(\mathbf{r}) {\rm d}^2\mathbf{r} =\pi r_0^2$.
Introducing the dimensionless repulsion length $\ell = \xi/r_0$, one gets $E_{\rm rep}(a) =\pi^2 E_0 r_0^4 f(a)$ where 
\begin{equation}
f(a)=\frac{\ell^2}{\sqrt{\ell^2+2a^2}\sqrt{\ell^2+2/a^2}}
\end{equation}
\begin{figure}[t!]
\centering
\includegraphics[width=0.7\columnwidth]{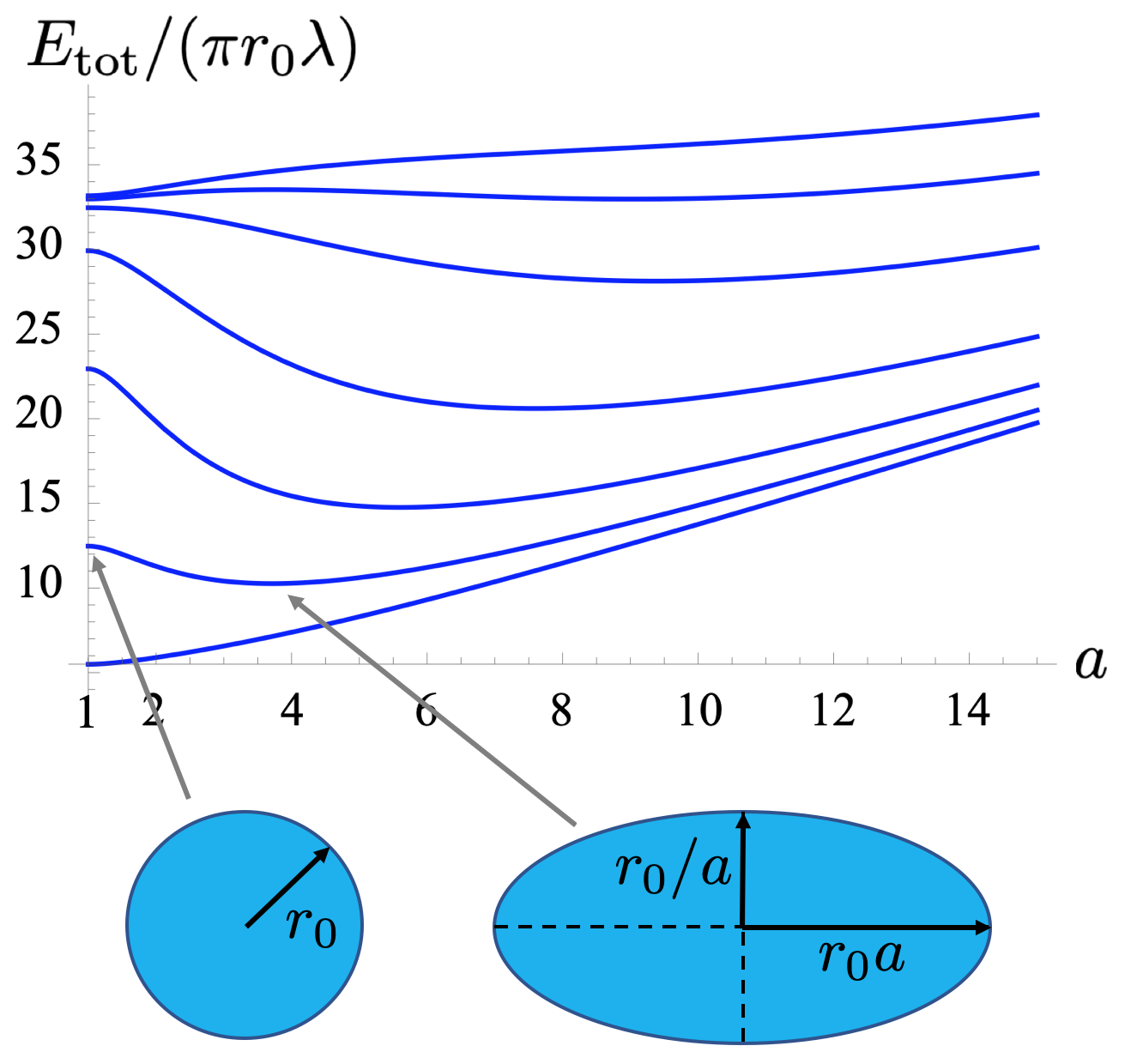}
\caption{The nanodomain adimensional energy $E_{\rm tot}/(\pi r_0\lambda)$ as a function of the ellipse aspect ratio parameter $a$, for $\varepsilon = 10$ and $\ell=0.5$, 1, 2, 4, 8, 12 and 16, from bottom to top.}
\label{fig:Etot}
\end{figure}
The line energy reads $E_{\rm line} = \lambda P$ where $\lambda$ is the line tension. The circumference $P$ of an ellipse of semi-axes $\alpha$ and $\beta$ is given by the elliptic function. However, a very good approximation by Ramanujan is
$P \simeq \pi \left[ 3(\alpha+\beta) - \sqrt{(3\alpha + \beta)(\alpha+3\beta)}   \right]$, thus $E_{\rm line}(a) = \pi \lambda r_0 g(a)$ with 
\begin{equation}
g(a)=\left[ 3\left(a+\frac1{a} \right) - \sqrt{\left(3a + \frac1{a}\right)\left(a+\frac3{a}\right)}   \right].
\end{equation}

\fig{fig:Etot} shows how $E_{\rm tot}/(\pi r_0\lambda)= \pi \varepsilon f(a)+g(a)$ behaves in function of $a$ for various values of $\ell$, where we have introduced the new dimensionless parameter 
\begin{equation}
\varepsilon = \frac{E_0 r_0^3}{\lambda}
\end{equation}
measuring the relative strengths of repulsion and line energies. One observes that there is a range of values of $\ell$ for which $a=1$ is not an energy minimum.
It implies that there exist values of the repulsion length $\xi$ for which the most stable nanodomain shape is an ellipse ($a>1$), and not a disc ($a=1$). 
Their relative stability can be addressed by examining the behavior of $E_{\rm tot}$ close to $a=1$. Expanding $f$ and $g$ at order 2 reads~\footnote{Using the exact elliptic function leads to the same expansion at order 2 for $g$.} $f(a)  \simeq f(1)  - \frac{4 \ell^2}{(2+\ell^2)^3}    (a-1)^2 $ and $g(a)  \simeq  g(1) +   \frac32  (a-1)^2 $.
Introducing $A_\ell =4 \ell^2/(2+\ell^2)^3$, it follows that 
$\frac12 \frac{{\rm d}^2 E_{\rm tot}}{{\rm d}  a^2} =  - \pi^2 E_0 r_0^4A_\ell + \frac32 \pi \lambda r_0$. Ellipses are stable when $\frac{{\rm d}^2 E_{\rm tot}}{{\rm d}  a^2} <0$, i.e. $A_\ell > \frac3{2 \pi \varepsilon}$; $A_\ell$ has a maximum $A^*=\frac8{27}$ at $\ell^*= 2$. There exists a region of stability of ellipses iff $A^* > 3/(2 \pi \varepsilon)$, i.e. $\varepsilon > \frac{81}{16 \pi}$.

Since the numerical prefactors in our expressions come from the choices of repulsive potential Gaussian shape $\varphi$ and Gaussian density profiles in the nanodomain, we simplify the principal results as follows: (i)~Elliptic domains are stable for $E_0 r_0^3>\lambda$, i.e. for  large enough domain radius $r_0$, strong enough repulsion strength $E_0$ or weak enough line tension $\lambda$. (ii)~If this condition is satisfied, $\ell=\xi/r_0$ must belong to an interval $[\ell^*_i,\ell^*_s]$ (see Fig.~S8) distributed around the maximum abscissa $\ell^*$ of order unity to stabilize ellipses with respect to discs. If the repulsion range is too short as compared to the domain radius, repulsion is not strong enough to destabilize circular domains. If it is too long, the gain in elongating the domain cannot compensate the line energy cost.  (iii)~The stability of ellipses requires that the repulsion range $\xi$ and the cluster radius $r_0$ are on the same order of magnitude.

In the context of protein nanodomains, $E_0 \approx \sigma C_0^2$ and we use the realistic values $\sigma=10^{-4}$~J/m$^2$, $C_0=0.05$~nm$^{-1}$, $\lambda=1$~pN~\cite{review}, and  $r_0=200$~nm (see below). Then $\varepsilon=E_0 r_0^3/\lambda > 1$ and $\pi r_0\lambda\simeq 150 k_BT$ (where $k_BT$ is the thermal energy at room temperature)~\footnote{For a small range of values of $\ell\gtrsim\ell^*_s$, two local minima of $E_{\rm tot}$ coexist, at $a=1$ and $a>1$, for example for $\ell = 12$ in \fig{fig:Etot}. Since minima are well pronounced with these values (the energetic barrier is $\gg k_BT$), one may find coexistence of discs and ellipses.}. Hence our scaling law shows that for realistic parameter values, circular domains can become unstable even at the sub-micrometric scale in the cell membrane context.
It is difficult to go beyond this rough estimate because the lipid and protein composition of the domains in which the receptors evolve is largely unknown at the current stage of knowledge. Thus the parameters $\lambda$ and $E_0$ cannot be precisely quantified.

\subsection*{Numerical simulations confirm that domains on a vesicle elongate when curving species concentration increases}
\begin{figure}[t!]
\centering
\includegraphics[width=0.9\columnwidth]{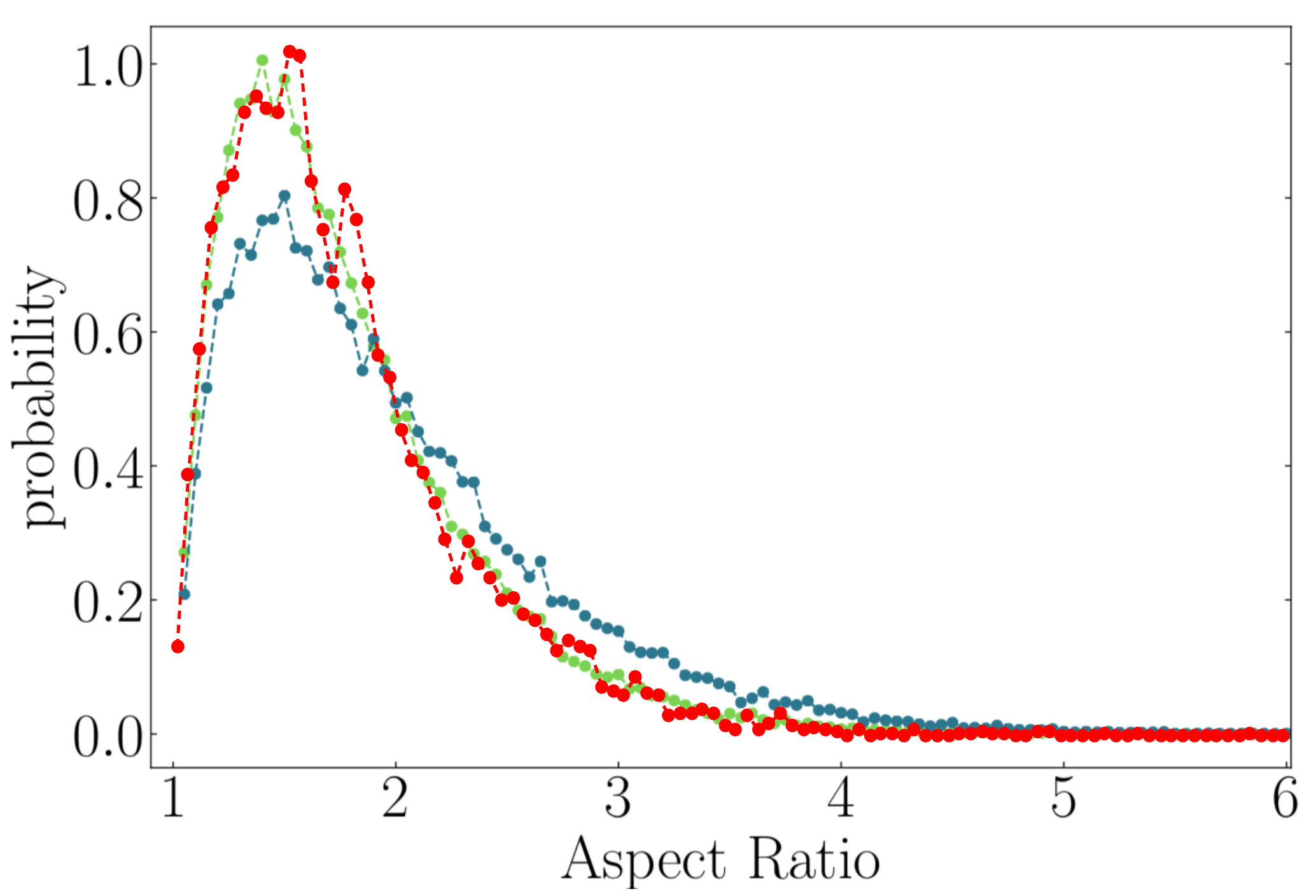}
\includegraphics[width=0.9\columnwidth]{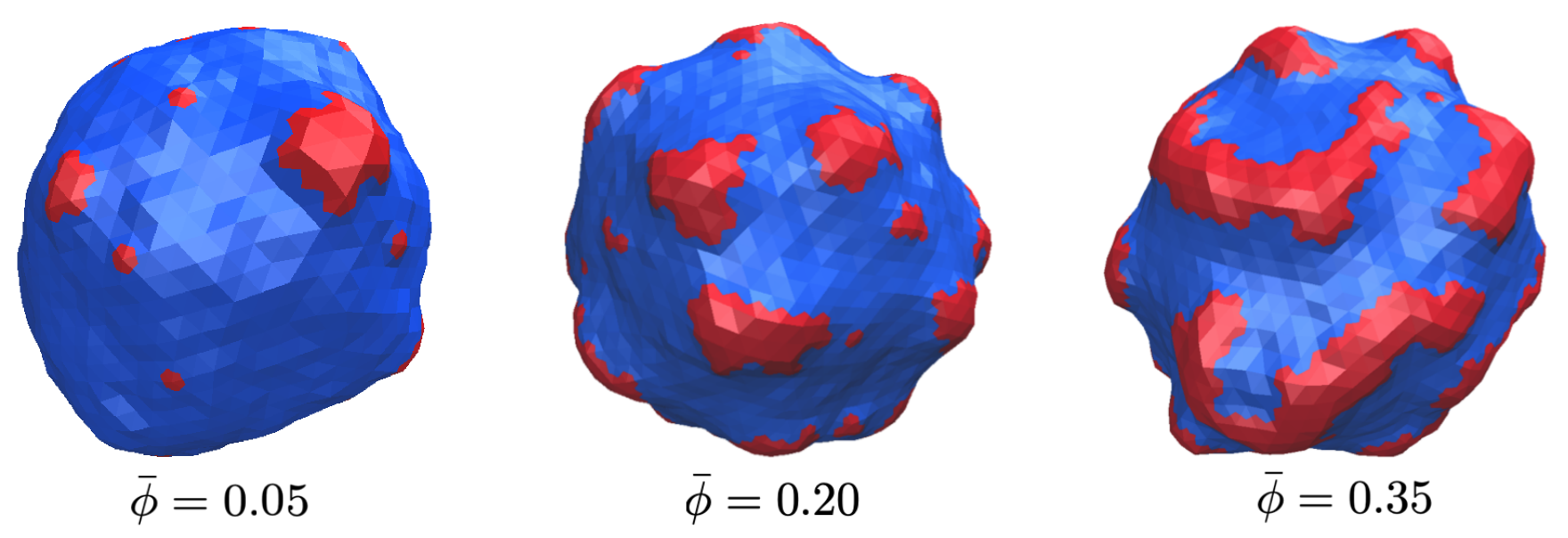}
\caption{Top: Simulation aspect ratio probability distributions of A (red) domains for vesicles with $\bar{\phi}=0.05$ (red), $\bar{\phi}=0.20$ (green) and $\bar{\phi}=0.35$ (blue). Other parameters are $c_1=8.0$, $\tilde{\sigma}=300$ and $\tilde{J}_I=0.5$. Whereas the  $\bar \phi =0.05$ and $\bar \phi =0.2$ distributions are close, the $\bar \phi = 0.35$ one is significantly different from the $\bar \phi = 0.20$ one ($p$-value below the computer accuracy, KS statistical test, see SM).  Bottom: simulations snapshots of the corresponding vesicles with the given values of $\bar{\phi}$.}
\label{fig:Etot2}
\end{figure}

To confirm and illustrate these results, we perform MC simulations, where we also observe elongated nanodomains, in particular when increasing concentration of the component forming the domains. We use a vesicle model, the discretized version of a continuous biphasic membrane model, developed in Ref.~\cite{Cornet2020} (see also SM): a lattice-gas model, with Ising interaction parameter $J_I$, describes the binary mixture of two species A and B. It is coupled to a discretized Helfrich model accounting for the membrane elasticity,  
where the local spontaneous curvature $C$ depends on the composition. In these simulations, the species A can be considered as a phase containing the membrane proteins of interest and/or particular lipids. It has a spontaneous curvature $C_{\rm A}$ different from the majority phase (species B with $C_{\rm B}=2/R<C_{\rm A}$ where $R$ is the average radius of the vesicle).  We chose  $c_1=R(C_{\rm A}-C_{\rm B})=8$~\footnote{The range of parameters of interest is restricted: we want to study rather small and numerous domains, which implies that $c_1=R(C_{\rm A}-C_{\rm B})$ has to be large enough~\cite{Cornet2020}. On the other hand, if $c_1$ is too large the domains get as small as the lattice spacing and their AR cannot be determined accurately.}. Besides, we observe in the experimental domains as shown in \fig{fig:traj} that  boundary fluctuations are small. This implies that the line tension $\lambda$ of the domain boundary has to be high enough and thus that the interaction parameter $\tilde{J}_I\equiv J_I/(k_BT)$ is significantly larger than its critical value $\tilde{J}_{I,c}=\ln(3)/4\simeq 0.27$ for an hexagonal lattice. Therefore we focus on the value  $\tilde{J}_I=0.5$.  We then run simulations with a typical value $\kappa=20~k_BT$ and a dimensionless surface tension $\tilde{\sigma}\equiv\sigma R^2/(k_BT)=300$, which corresponds to quasi-spherical vesicles~\cite{GG2}, and study rather low A-species concentration $\bar{\phi}=0.05$ and 0.20 versus a higher one $\bar{\phi}=0.35$. We run long simulations, up to $3\times10^{10}$ MC steps on 2562 vertices to have good enough statistical sampling~\cite{Cornet2020}.
To measure the AR of domains lying on a quasi-spherical surface, we project each of them onto the plane tangent to the average sphere at the domain center of mass and to compute domain covariance matrix with its in-plane coordinates $(x,y)$. The AR is then simply the ratio of the square roots of its two eigenvalues (see SM).
\fig{fig:Etot2} shows obtained AR distributions. 
We note that the curves intersect at $\mathrm{AR}_0\simeq 2$. Domains with an $\mathrm{AR}\leq2$ (respectively $\mathrm{AR}>2$) are thus called roundish (resp. elongated). The increase of concentration does not have any significant effet on the AR distribution at low concentrations, between $\bar \phi=0.05$ and 0.20. By contrast, when $\bar \phi$ grows from 0.20 to 0.35, one observes an increase in the proportion of elongated domains from $28\%$ to $40\%$. In the SM, we also measure the typical cluster sizes. 
As expected~\cite{Cornet2020}, the domains for $\bar{\phi}=0.35$ have a larger typical size than the ones at $\bar{\phi}=0.2$.
The average cluster size growing when $\bar{\phi}$ is increased, more clusters fulfil the condition $r_0^3>\lambda/E_0$ and therefore become elongated as predicted by the analytical model. 
Now we estimate the parameter $\varepsilon$. We get $\lambda \sim 0.01$~pN from the value of $J_I$ and a vesicle radius $R=10$~$\mu$m~\footnote{In Ref.~\cite{TheseJulie}, it is explained how renormalization group methods allow one to relate $\lambda$ to $R$ and $J_I$. In fact, we chose a value of the vesicle radius $R=10$~$\mu$m for illustration sake, but the dimensionless value of $\varepsilon$ does not depend significantly on this value owing to the way the different parameters entering $\varepsilon=\sigma C_0^2 r_0^3/\lambda$ scale with $R$: $\sigma \propto R^{-2}$, $C_0 \propto R^{-1}$, $r_0 \propto R$ and $\lambda \propto R^{-1}$.}. The surface tension has been estimated to be $\sigma \sim 10^{-8}$~J/m$^2$ with those parameters and the domain curvature is $C_{\rm A} = (c_1+2)/R \sim 1$~$\mu$m$^{-1}$~\cite{TheseJulie}. The observed domain radius $r_0$ is about $2$~$\mu$m. It follows that $\varepsilon={E_0 r_0^3}/{\lambda}  \approx 10$, consistently larger than 1, and $\pi r_0\lambda\simeq 30k_BT$. In the SM, we are led to similar conclusions by exploring a second numerical model where proteins are represented as point-like objets~\cite{Destainville2008} (Figs.~S10 and~S13).

\subsection*{HIV receptor nano-domains tend to elongate under overexpression}

Our laboratory has long been interested in the early mechanisms of HIV infection and in the role of membrane domains in this process. To study the influence of protein overexpression on the shape of the domains, we have chosen to take advantage of the 293-Affinofile cell lines developed by Johnston {\em et al.}~\cite{Johnston2009}. This inducible cell line was originally engineered to manipulate the CD4 and CCR5 expression levels over a range covering that found on primary HIV-1 target cells. The proteins CD4 and CCR5 can be simultaneously and independently regulated with the help of variable concentrations of minocycline and ponasterone, respectively (Fig.~S1).
We have stably transfected this cell line to express CXCR4, the third membrane protein that can also be involved in HIV infection. We have established two stable cell lines, one expressing a low number ($15 000 \pm 2000$ per cell) of CXCR4 and one expressing a high number ($120 000 \pm 7000$) of this protein (Fig.~S2). Thanks to this, we were able to generate cells presenting any possible combination of protein expression at their surface (Table~\ref{Table1}, Figs.~S1 and S2).
Note that this model is relevant regarding HIV infection since, whatever the expression level of the three proteins, the cells can be infected by HIV-1 viruses (see Fig.~S3)~\cite{Colin2018,Armani2021}.

We have performed SPT experiments to track these different proteins in different conditions. The collected data give the positions of the tracked proteins every 40~ms for  $\geq 30$~s. 544 individual trajectories have been acquired with different expression levels of each protein. 
35\% of the trajectories showed a free diffusion over the entire duration of the measurement and 65\% showed confined diffusion (either permanently or transiently).
For each of the latter trajectories, we have isolated the confinement zones thanks to the confinement index $\Lambda$ of Ref.~\cite{Meilhac2006} (see Materials and Methods). The duration of the confined trajectories is at least 4 times larger, and in general much longer, than the diffusion time to explore the whole confinement zone (Fig.~S4). Indeed by construction, the confinement index detects a transient confinement zone because the spatial extent of the trajectory is significantly smaller than it would be expected for free diffusion. We have then measured the sizes and shapes of these confinement zones. If we again define the radius $r_0$ through the ellipse area $A=\pi r_0^2$ (see analytical model), we find the typical values $r_0\simeq150$~nm (respectively 200~nm)  in the low (resp. high) expression state (Table~S1). 
\begin{figure}[t!]
\centering
\includegraphics[width=0.9\columnwidth]{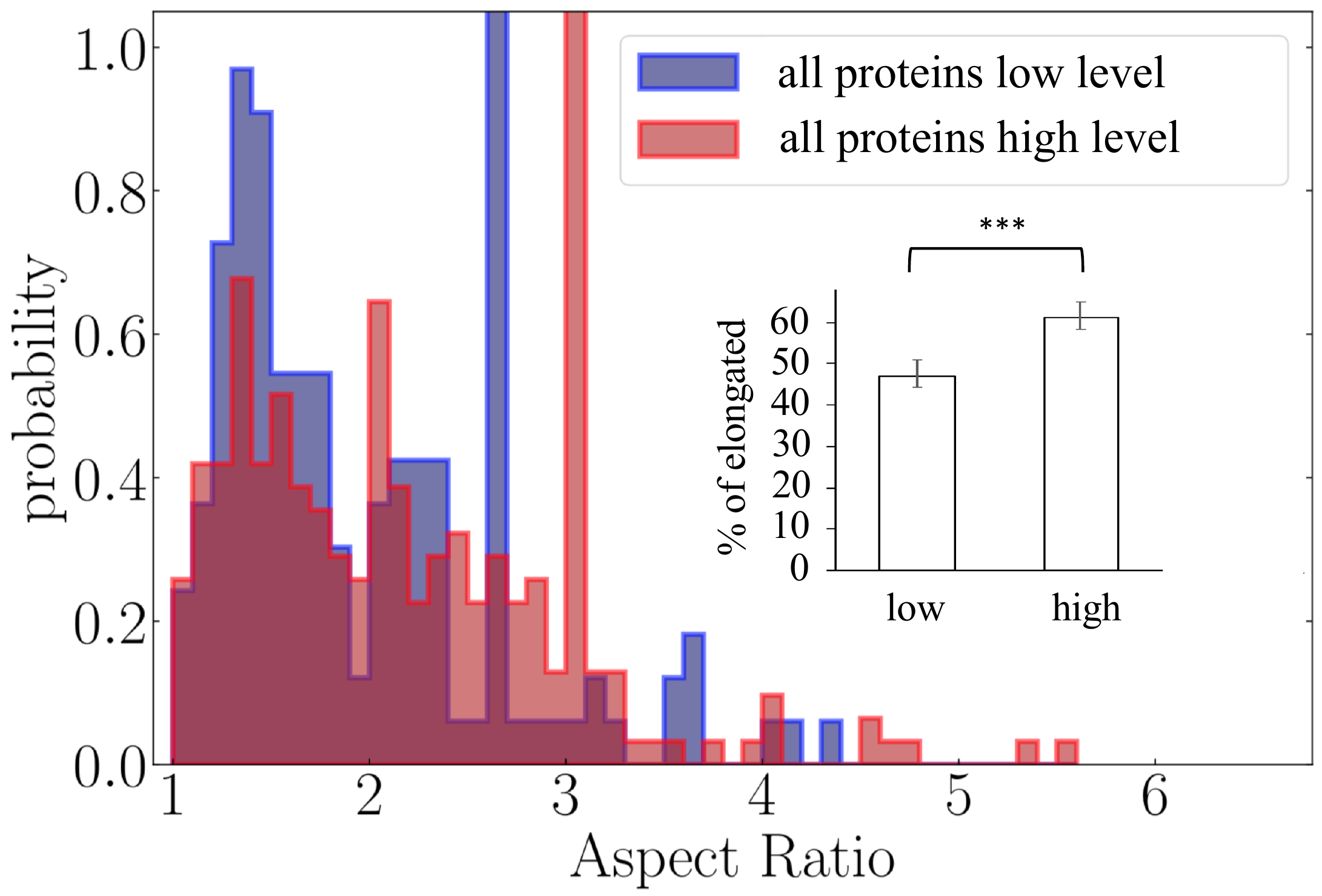}
\caption{Experimental aspect ratio probability distributions when all proteins have a low (blue, $n_{\rm low}=165$ measurements in 113 cells acquired in 21 independent experiments) or high (red, $n_{\rm high}=317$ measurements in 211 cells acquired in 38 independent experiments) expression level ($p$-value $<10^{-7}$ for KS statistical test, see SM). 
Inset: fractions of elongated nanodomains in both conditions. Error bars are standard errors of the mean.}
\label{fig:Etot3}
\end{figure}

\begin{figure}[h!]
\centering
\includegraphics[width=5.7cm]{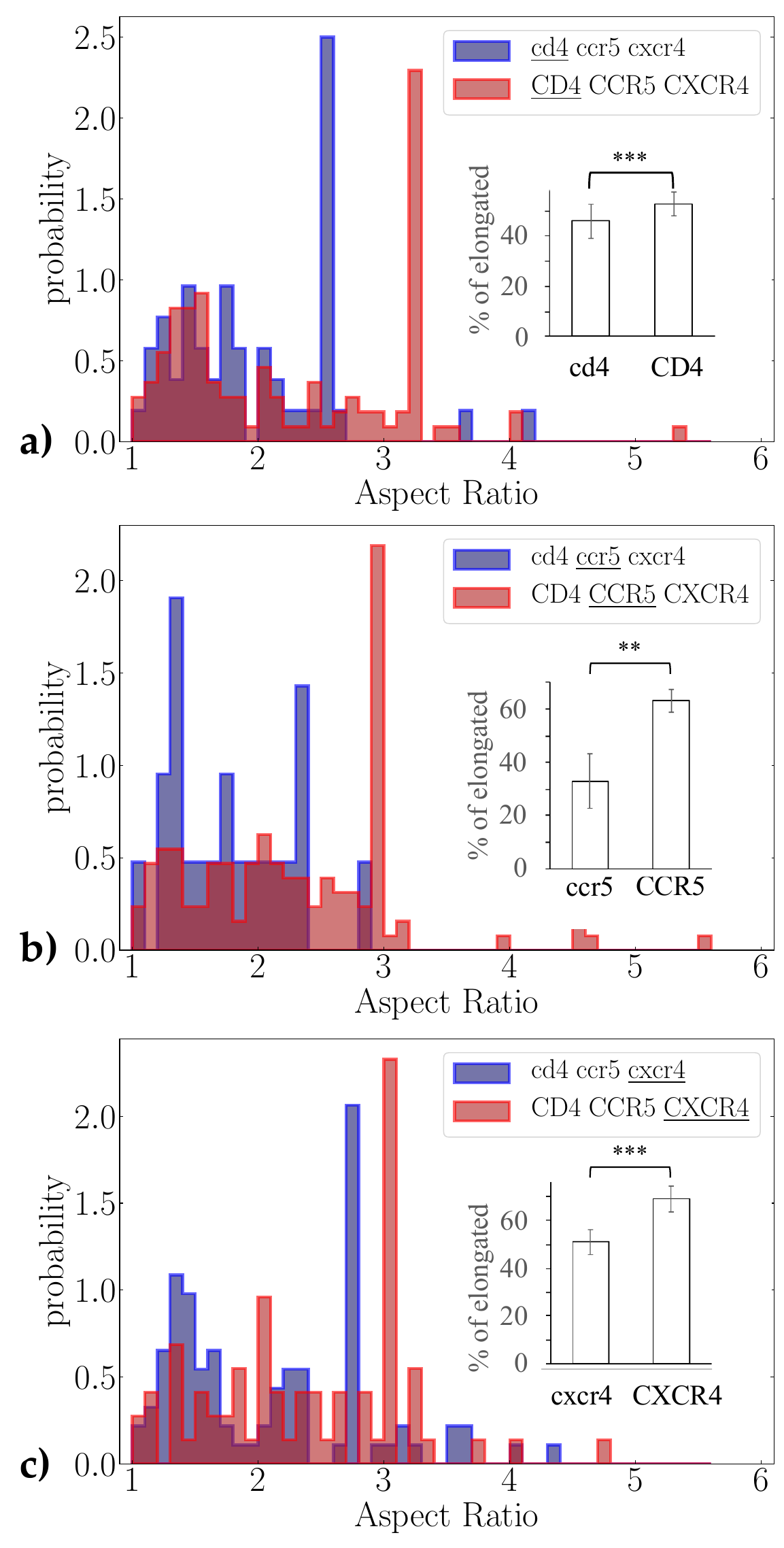} 
\includegraphics[width=5.7cm]{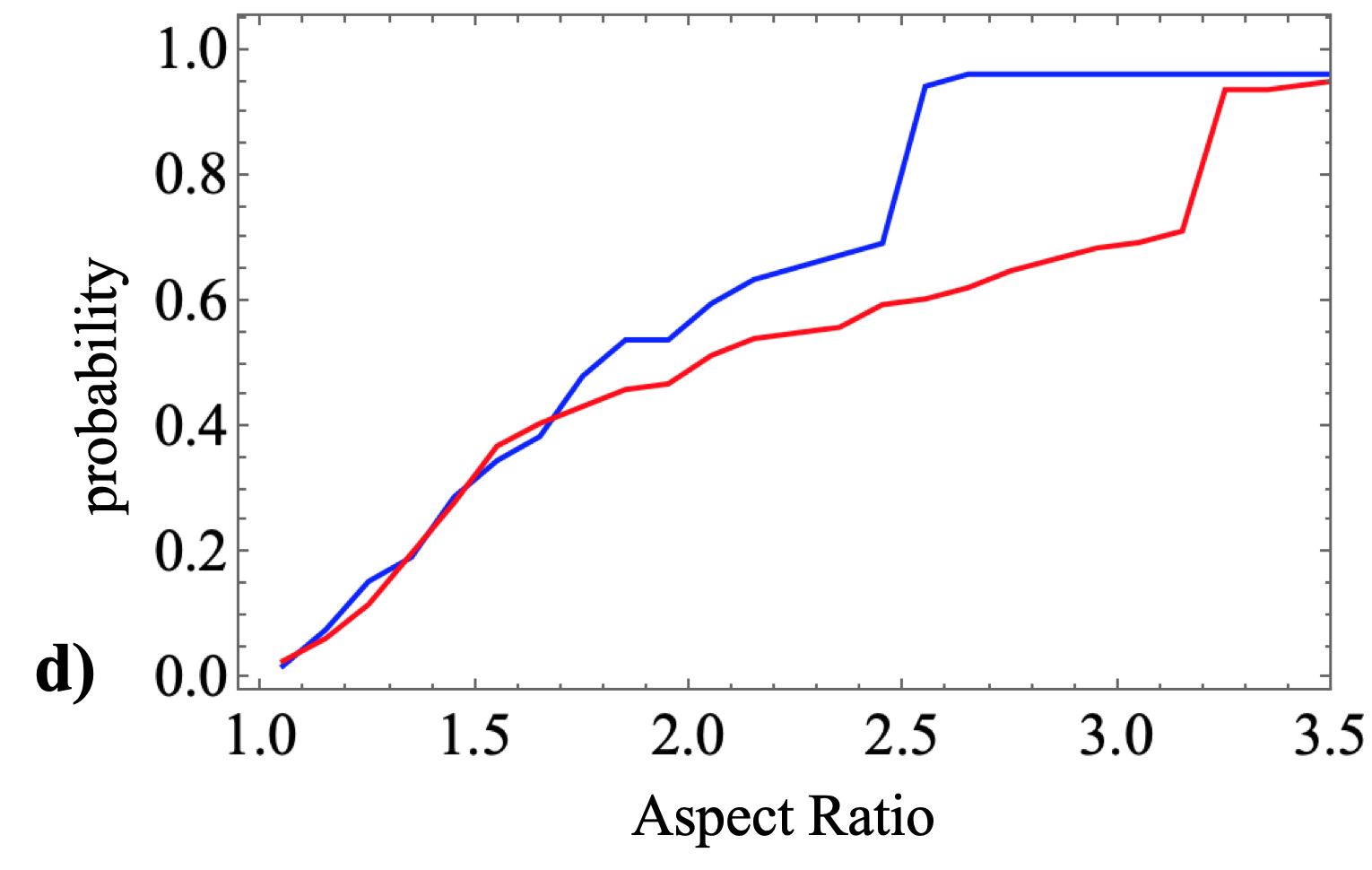} 
\caption{Same as \fig{fig:Etot3} for the three types of proteins tracked separately, when they all have low (blue) and high (red) expression levels. (a) $p \simeq 2 \times 10^{-4}$, $n_{\rm low}=52$ measurements in 29 cells acquired in 7 independent experiments, $n_{\rm high}=111$ measurements in 83 cells acquired in 13 independent experiments, (b) $p\simeq 0.002$, $n_{\rm low}=21$ measurements in 15 cells acquired in 4 independent experiments, $n_{\rm high}=131$ measurements in 88 cells acquired in 15 independent experiments, and (c) $p\simeq 0.001$, $n_{\rm low}=92$ measurements in 69 cells acquired in 10 independent experiments, $n_{\rm high}=75$ measurements in 47 cells acquired in 9 independent experiments. Protein name is written in lowercase letters when expressed at low level and in uppercase letters when overexpressed. The protein that has been tracked is underlined. Error bars on histograms are standard errors of the mean (s.e.m.). The $p$-values are not calculated with these error bars but with the full distributions via a Kolmogorov-Smirnov statistical test (see below). In (d), the cumulative distribution of (a).
\label{fig:exp1}}
\end{figure}

In \fig{fig:Etot3} we compare the AR distributions of the three proteins that we have pooled in the context where the three proteins have simultaneously either low or high expression levels, one of them being tracked (the individual distributions before pooling are shown in \fig{fig:exp1}). We observe a significant increase in the proportion of elongated versus rounded domains.
In the SM, we also explain that some confinement zones with $\mathrm{AR}\leq2$ are however classified as elongated because they are in fact curled or coiled elongated nanodomains. They appear as peaks in \fig{fig:Etot3}, to be compared to Fig.~S6 (where the distribution without curled or coiled nanodomains is also shown). In overexpressed conditions, the AR distribution is significantly shifted towards higher values and the proportion of elongated nanodomains raises from 47\% to 61\%.

\begin{figure}[t]
\begin{center}
\includegraphics[width=5.7cm]{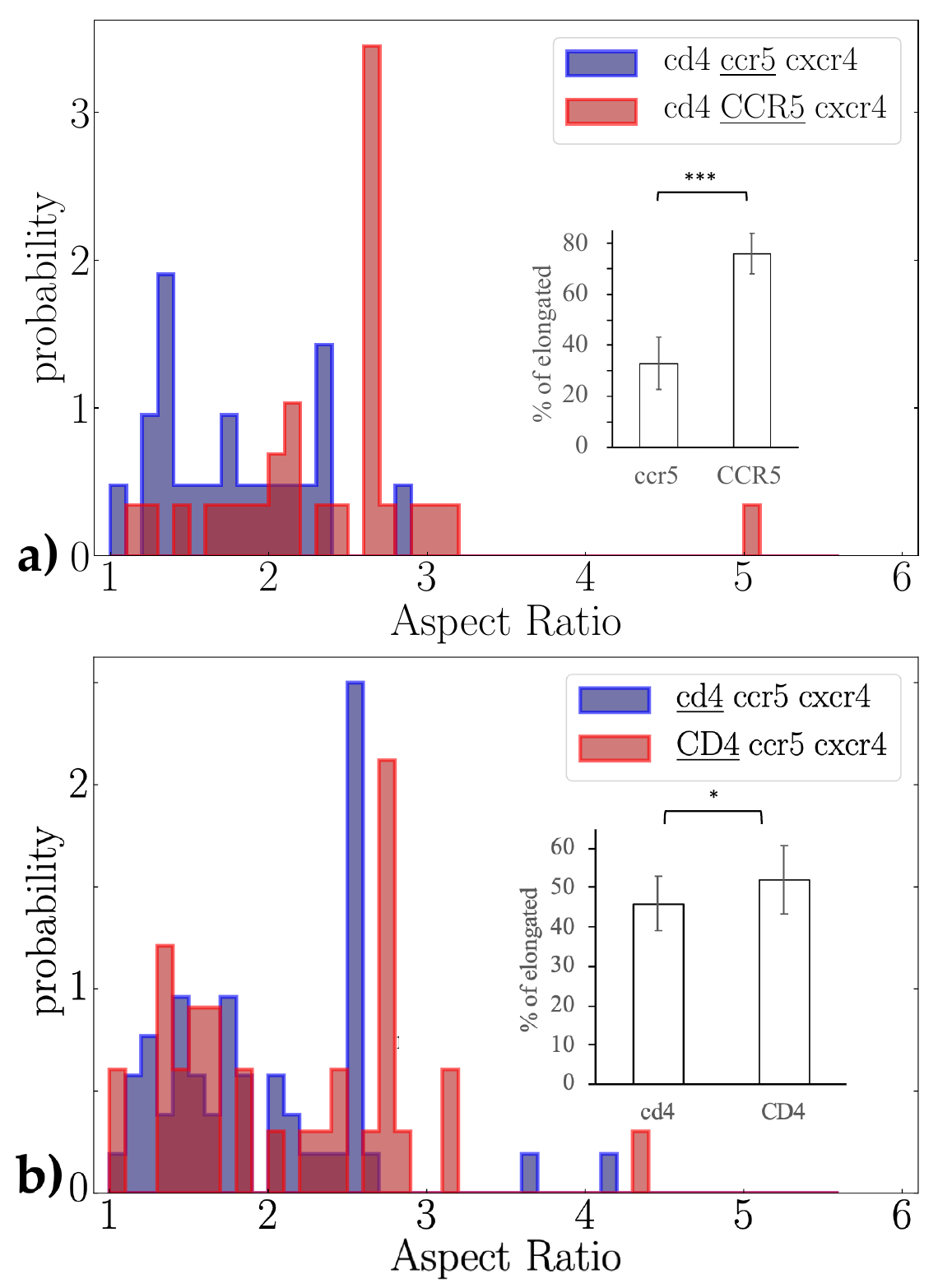}
\end{center}
\caption{Same as \fig{fig:Etot3}  for different conditions [(a) $p\approx9\times10^{-4}$, $n_{\rm low}$, $n_{\rm high}=29$ measurements in 16 cells acquired in 5 independent experiments, and  (b)~$p\approx 0.03 $, $n_{\rm low}$ and $n_{\rm high}=33$ measurements in 19 cells acquired in 6 independent experiments. Protein name is written in lowercase letters when expressed at low level and in uppercase letters when overexpressed. The protein that has been tracked is underlined.}
\label{fig:exp2}
\end{figure}
Then, we have compared the behavior of each protein individually in a low or high expression context.
In the Figures displayed below, we use the following notations for the experimental conditions: when a protein is expressed at a low expression level, its name is written in lowercase letters, when a protein is overexpressed, its name is written in uppercase letters. The protein that has been tracked is underlined.
We also observe a significant increase in the proportion of elongated versus rounded domains regardless of the protein being considered (\fig{fig:exp1}). Alternatively, the cumulative distribution of panel~(a) shown in panel~(d) displays a shift to higher AR values upon overexpression.
To go further, we have focused on the single-spanning CD4 and seven-spanning CCR5 receptors that have been abundantly studied in the literature~\cite{Gaibelet2006,Dumas2014,Beauparlant2017,Shaik2019}. We have analyzed changes of the shape of the nanodomains of each of these proteins when it is the only one to be overexpressed. In \fig{fig:exp2}, we observe that overexpression of the sole CCR5 is accompanied by a strong increase of the proportion of elongated nanodomains (33\% to 76\%) while for the CD4, we have observed a slight increase of the proportion of elongated nanodomains (46\% to 52\%), however statistically significant. 
 
\section*{Discussion}

This work thus combines theoretical and experimental approaches to show that increasing the concentration of the minority phase can lead to a noticeable elongation of nanodomains in membranes, as already observed experimentally~\cite{Merklinger2017} or in numerical simulations~\cite{Konyakhina2011}. However, to our knowledge, this effect had never been quantified so far. 
A simple physical mechanism in thermodynamic equilibrium can be put forward to explain why elongated nanodomains are more stable than roundish ones under favorable circumstances. When $\bar\phi$ grows, nanodomains become more and more numerous with a growing typical size $r_0$ (see Fig.~S11)~\cite{Foret2008,Cornet2020}. However, too large domains are intrinsically unstable, because of the effective long-range repulsion due,  for example, to membrane deformation induced by the spontaneous curvature of the domain constituents. One way of dealing with this instability is to generate more elongated nanodomains above a critical size, in which the repulsive energy (of magnitude $E_0$) is lower, at the price, however, of a higher line energy, proportional to the line tension $\lambda$. More quantitatively, we propose a scaling argument, which writes $E_0 r_0^3>\lambda$, predicting when circular domains become unstable at the benefit of elongated ones, provided that the typical domain size $r_0$ must be comparable to the range $\xi$ of the repulsion. In our SPT experiments, we have measured $r_0\approx 200$~nm when receptors are overexpressed which indeed corresponds to the realistic value $\xi=\sqrt{\kappa/\sigma}\sim 100$~nm in cells~\cite{review}. 

Pointing out the analogy between experimental and numerical nanodomain morphologies implicitly assumes that a curvature-composition coupling mechanism is at play in the case of HIV receptror-containing nano-domains, at least for CCR5 and CXCR4, two class-A G-Protein Coupled Receptors (GPCRs). 
The spontaneous curvature induced by class-A GPCRs has very recently been investigated in detail in live cells~\cite{Rosholm2017}. A spontaneous curvature of about 0.04~nm$^{-1}$ is deduced, presumably related to the crystal structures of those GPCRs that reveal their transmembrane part to be up-down asymmetric across the bilayer. 
This spontaneous curvature is typically in the range of values that can promote sub-micrometric nanodomains, as expected~\cite{review}. 

To our knowledge, no such measurements have been performed on CCR5 or CXCR4. But they belong to the same class A~\cite{Lodowski2009} and thus share structural similarities with those of Ref.~\cite{Rosholm2017}. 
The effect of overexpression is less marked for CD4. CD4 has only one transmembrane segment and its impact on the local membrane curvature should be less important than for CCR5 which has seven transmembrane segments.

The present work then shows that the accumulation of proteins into nanodomains can conduct to a change of their morphology. This could probably be extrapolated to other membrane proteins since such ``untypical'' nanodomain shape has already been observed with other proteins in different cell types without being explained so far. However, this mechanism should not be confused with the one studied for example in Ref.~\cite{Noguchi2022}. There, very anisotropic proteins adsorbed on the membrane, such as BAR domains, tend to organize in linear structures with a strong order due to strong interactions. In these structures, each protein has a fixed position and cannot diffuse easily, contrary to what we observe here. In our case, the domains are likely much more disordered, with a liquid-like order.  Here the proteins under study may also induce a moderate anisotropic curvature, but anisotropy is rapidly averaged out due to thermal fluctuations and fast rotational diffusion~\cite{Meilhac2011}.

Domain elongation thus reveals local accumulation of specific proteins in the cell membrane. Such an effect, influencing biological processes as HIV entry~\cite{Kuhmann2000,Baker2007}, can be revealed by SPT thanks to its unique performances. From a soft-matter physics viewpoint~\cite{Seul1995}, we conjecture further that elongation of nanodomains is the signature of the transition between 2D hexagonal and lamellar phases, where the minority phase transforms into parallel stripes~\cite{Sear1999}. Indeed, in phase diagrams ensuing from approximate calculations, a region of coexistence between these two phases was identified~\cite{Harden2005} that might contain the elongated nanodomain stability region, as numerical simulations suggest it~\cite{Konyakhina2011,Mukhopadhyay2008}. Refined calculations will be necessary to get a full understanding of meso-patterning  in the future.

\section*{Acknowledgments}

We are indebted to Blandine Doligez for her sound advice on non-parametric statistical tests and to Thorsten Lang for indicating us the existence of elongated nanodomains observed by STED under overexpression of syntaxin. We thank Samuel Tranier for his help in the determination of the diameter of CD4, CCR5 and CXCR4 proteins.

\newpage
\hspace{15cm}



\section*{Supplementary material}

# Supplementary Material
# Protein overexpression can induce an elongation of cell membrane domains

Julie Cornet[†], Pascal Preira[*], Laurence Salomé[*], Frédéric Daumas[*],
Bernard Lagane[‡], Nicolas Destainville[†], Manoel Manghi[†], Fabrice Dumas[*]

[†] *Laboratoire de Physique Théorique, Université Toulouse III, CNRS, 31062 Toulouse, France*
[*] *Institut de Pharmacologie et de Biologie Structurale,
Université Toulouse III, CNRS, 31062 Toulouse, France*
[‡] *Infinity, Université de Toulouse, CNRS, Inserm, Toulouse, France*

(Dated: December 2, 2022)

## I. CELL LINES

To study HIV infection processes, Johnston et al. [1] have developed an inducible cell line in which two membrane proteins (CD4 and CCR5) can be simultaneously and independently regulated with a large range of surface expression with the help of variable concentrations of minocycline and ponasterone, respectively (Figure S1). We have stably transfected this cell line to express the third membrane protein (CXCR4) that can be involved in HIV infection. We have established two stable cell lines, one expressing a low number ($15000 \pm 2000$ per cell) of CXCR4 and one expressing a high number ($120000 \pm 7000$) of this protein (Figure S2). The CXCR4 copy number of each line was determined in triplicate using Quantibrite system described in the Mat. and Meth. section of the main text. The average of the minimum and maximum values of CD4, CCR5 and CXCR4 copy numbers are reported in Table 1 of the main text.

Figure S3 shows that the cell lines that we have established can be infected by X4- and R5-viruses (NL 4-3 and Bx08 respectively).

Based on these figures, we can also calculate a rough estimate of the protein density (or surface ratio) occupied by the proteins on a typical cell membrane in the case of overexpression. As shown in Table S1, the membrane surface ratio accupied by proteins is less than 1%. However, one should consider that a real cell is a far more complex system. The proteins do not cluster alone but also recruit certain lipids and partner proteins so that the surface ratio of the concerned domains is probably significantly higher than the value obtained by considering the sole tracked proteins.

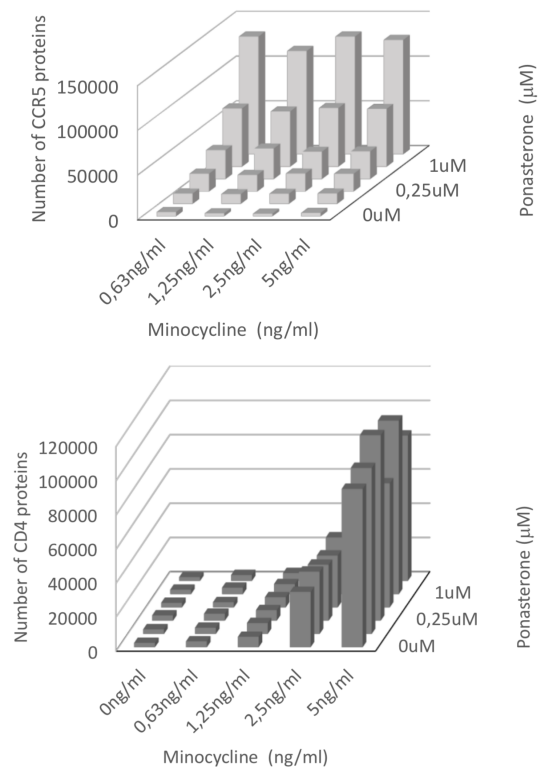

FIG. S1: CD4 and CCR5 can be independently and simultaneously regulated. Top: ponasterone induces the expression of CD4 proteins on the same manner whatever the minocycline concentration. Bottom: Minocycline induces the expression of CCR5 proteins on the same manner whatever the ponasterone concentration.

## II. TRANSIENTLY CONFINED TRAJECTORIES

In Figure S4 is shown the distribution of the ratio between the trajectory time and the typical diffusion time to explore the domain of size $L$ measured in the experiments, $L^2/(4D)$ (where $D$ is the measured lateral diffusion coefficient). For the trajectories studied, this ratio is always larger than 4, thus confirming that the full domain as indeed been explored. This ensures that elongated trajectories account for merely elongated domains and are not due to a bias arising from a too short acquisition time leading to the partial exploration of a domain.

## III. QUANTIFICATION OF MEMBRANE DOMAIN SHAPE

In SPT, some diffusion modes can give rise to *apparently* elongated domains. Notably, directed diffusion trajectories might be confused with confined trajectories in an elongated domain. We put directed diffusion trajectories aside through the characterization of their MSD [3].

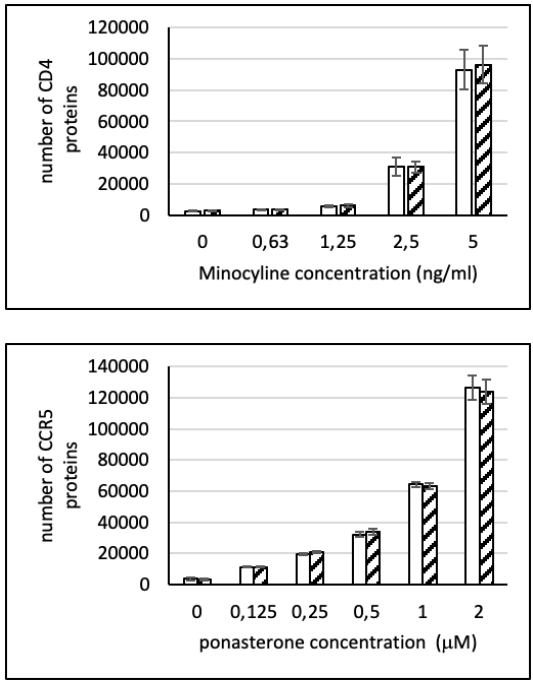

FIG. S2: The expression of CD4 and CCR5 can be controlled in the inducible 293-Affinofile cell lines that stably express high or low level of CXCR4 (dashed and white bars respectively). Minocycline (top) and ponasterone (bottom) induced CD4 and CCR5 expression, respectively, in a concentration dependent manner without affecting one another.

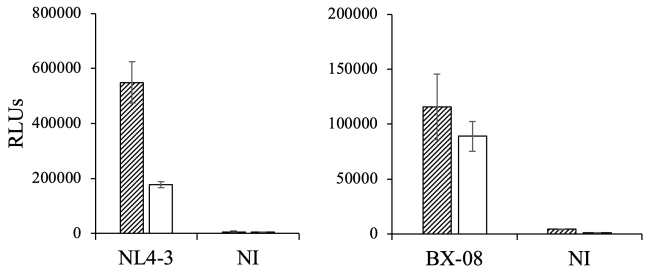

FIG. S3: Affinofile stable cells lines expressing high or low amounts of CXCR4 (dashed bars and white bars respectively) were induced in order to express CD4 and CCR5. They are successfully infected by R5 tropic (Bx08) or X4 tropic (NL4-3) viruses. Experiments have been carried out according to Ref. [2] as described in Section VI of SM. Results represent the luciferase activity in the cell lysates, expressed as relative light units (RLUs). Uninfected cells (NI) represent negative controls.

Note also that we did not observe any walking confined diffusion mode trajectories, where the tracked particle is confined in a domain that itself diffuses, which can also be confused with a trajectory in an elongated domain. Then we isolated the confined part of the trajectories thanks to the confinement index developed in Ref. [4] (see this article for more details about this detection method). The data collected give the $(x,y)$ coordinates of the con-

|        | % of cell surface occupied by proteins ||
|        | low expression | high expression |
|--------|----------------|-----------------|
| CCR5   | 0.0072 %       | 0.249 %         |
| CXCR4  | 0.0356 %       | 0.285 %         |
| CD4    | 0.0012 %       | 0.0509 %        |
| Total  | 0.044 %        | 0.585 %         |

TABLE S1: Percentage of the cell surface occupied by each protein expressed at high or low concentration. Based on the structures of CCR5 (PDB ID: 4MBS), CXCR4 (PDB ID: 3OE9) and CD4 (PDB ID: 2KLU), we have estimated the surface occupied by each of these proteins in the plane of the cell membrane to be 9, 10.75 and 2.3 nm$^2$ respectively. We have considered that HEK cells have a radius of 6 $\mu$m and taken into account the number of copies of each protein presented in Table 1 of the main text to calculate the percentage of cell surface occupied by each protein at low or high expression level.

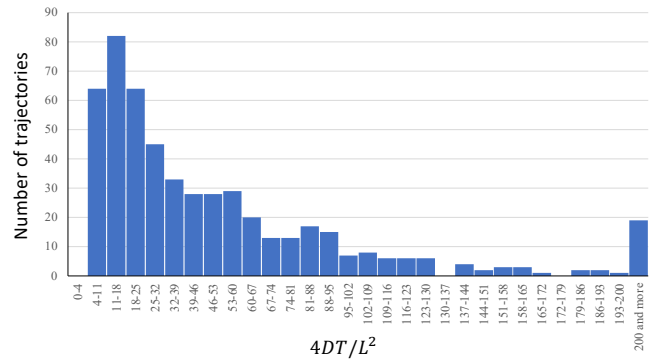

FIG. S4: Distribution of the ratio $4DT/L^2$ where $T$ is the confined trajectory duration of HIV receptors, $D$ their measured lateral diffusion coefficient, and $L$ the average size of the confinement zone.

fined parts of the trajectories of the tracked proteins every 40 ms. The localization precision has been measured to be of $\leq 7$ nm in both $x$ and $y$ directions by recording and analyzing 80 s videos of quantum dots immobilized in 15 % polyacrylamide gel [3, 5] (see Figure S5). Examples of trajectories are displayed in Figure 2 in the main text. Such trajectories clearly give us information about domain sizes and shapes, thanks to the unrivalled accuracy of SPT. To quantify size and shape, we determine the main axes of the observed domains, from the eigenvalues of their covariance matrix:

$$V = \begin{pmatrix} \sigma_x^2 & \text{cov}(x,y) \\ \text{cov}(x,y) & \sigma_y^2 \end{pmatrix} \quad (S1)$$

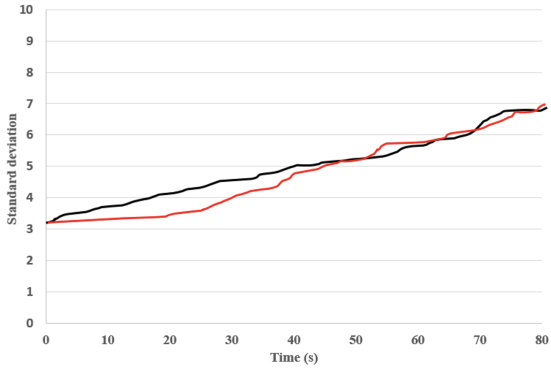

FIG. S5: Evolution of standard deviation of immobilized colloids as a function of time. To estimate the accuracy of our setup with respect to the determination of the particle position, we recorded a sequence of 80 s of several colloids deposited on a coverslip and immobilized in a 15% polyacrylamide gel. The standard deviations of the ≈ 30 colloids coordinates were found to increase with time indicating a limitation due to a thermomechanical instability of the microscope. After 80 s of measurements, the standard deviations $\sigma_x$ (red) and and $\sigma_y$ (black) were found to be equal to 7 nm.

where $\sigma_{x,y}^2$ are the variances and

$$\text{cov}(x,y) = \frac{1}{N}\sum_{i=1}^{N} x_i y_i - \left(\frac{1}{N}\sum_{i=1}^{N} x_i\right)\left(\frac{1}{N}\sum_{i=1}^{N} y_i\right), \quad (S2)$$

is the covariance for each confined trajectory. Let $\lambda_1 > \lambda_2$ be the real eigenvalues of the symmetric matrix $V$. We approximate the domain area by

$$\mathcal{A} \simeq 4\pi\sqrt{\lambda_1 \lambda_2} \quad (S3)$$

so that in the case where the domain is a disc in which trajectory points are uniformly distributed, one recovers the area of the disc. This also amounts to set that the semi-major and minor axes are defined by $r_1 = 2\sqrt{\lambda_1}$ and $r_2 = 2\sqrt{\lambda_2}$ when the domain is elliptic. Experimental values are given in Table S2.

In this Table we use the following notations for the experimental conditions: when a protein is expressed at a low expression level, its name is written in lowercase letters, when a protein is overexpressed, its name is written in uppercase letters. The protein that has been tracked is underlined.

The aspect ratio of the domain is then naturally defined as

$$\text{AR} = \sqrt{\frac{\lambda_1}{\lambda_2}} \quad (S4)$$

We set the threshold value above which domains are considered elongated to $\text{AR}_0 = 2$. It comes from the numerical distributions showing a "hinge point" for this value when concentration is increased and domains start to become more elongated, as observed in Fig. 5 in the main

|  | semi-minor axis (nm) | semi-major axis (nm) |
|---|---|---|
| <u>cd4</u> ccr5 cxcr4 | $120 \pm 14$ | $204 \pm 24$ |
| <u>CD4</u> CCR5 CXCR4 | $156 \pm 10$ | $280 \pm 18$ |
| cd4 <u>ccr5</u> cxcr4 | $127 \pm 13$ | $218 \pm 26$ |
| CD4 <u>CCR5</u> CXCR4 | $150 \pm 8$ | $288 \pm 14$ |
| cd4 ccr5 <u>cxcr4</u> | $130 \pm 8$ | $236 \pm 14$ |
| CD4 CCR5 <u>CXCR4</u> | $146 \pm 12$ | $312 \pm 32$ |

TABLE S2: Average semi-major ($r_1$) and minor ($r_2$) axes in various experimental conditions as inferred from the eigenvalues of the covariance matrix (see main text for the convention on experimental condition notations). Error bars are standard errors on means (s.e.m.). See text for notations.

text and Fig. S10 below.

In Figure S6 we compare the aspect ratio distributions of the cumulated data of all the proteins having low expression level on the one hand, and the cumulated data of all the proteins overexpressed on the other hand. We can already notice that the domains in overexpressed conditions show an aspect ratio distribution shifted towards higher values.

However, we have also noticed that for some domains, the aspect ratio was not a sufficient parameter to characterize their shape. Indeed some ring-shaped, spiral-shaped, or horseshoe-shaped domains lead to AR $\simeq$ 1, whereas they are in fact coiled elongated domains (Figure S7). We cannot numerically discriminate these domains through our sole AR analysis. In the experimental trajectories however we noticed that these domains were rather numerous ($> 10\%$), in majority with a horseshoe shape, which motivated us to implement a second verification, carried out on the domains that are classified by their aspect ratio as roundish (AR $< 2$): we calculate the domain barycenter and divide the domain into regular angular sectors emerging from this barycenter. We then compute the number of points per sector, as illustrated in Figure S7 (top).

We are interested in characterizing the fact that some sectors might contain no or a few points, much less than in the other sectors, corresponding to a hole in the angular distributions. We use the criterion described in Figure S7 (bottom): after computing the mean number of points per sector, we set a threshold $t$ below which the domain is considered to have a hole if at least one of its sectors contains less points than this value $t$, and then considered as coiled elongated. We tried different threshold values and chose to focus on two of them, $t = 25\%$ and 30% of the average number of points per sector. The distributions given in the figures are computed with $t = 25\%$. We checked visually that it discriminated most of the hidden elongated domains and included very few real "roundish" domains in this category. The modifications generated by this correction were on the order of few percents on the percentages of elongated domains for

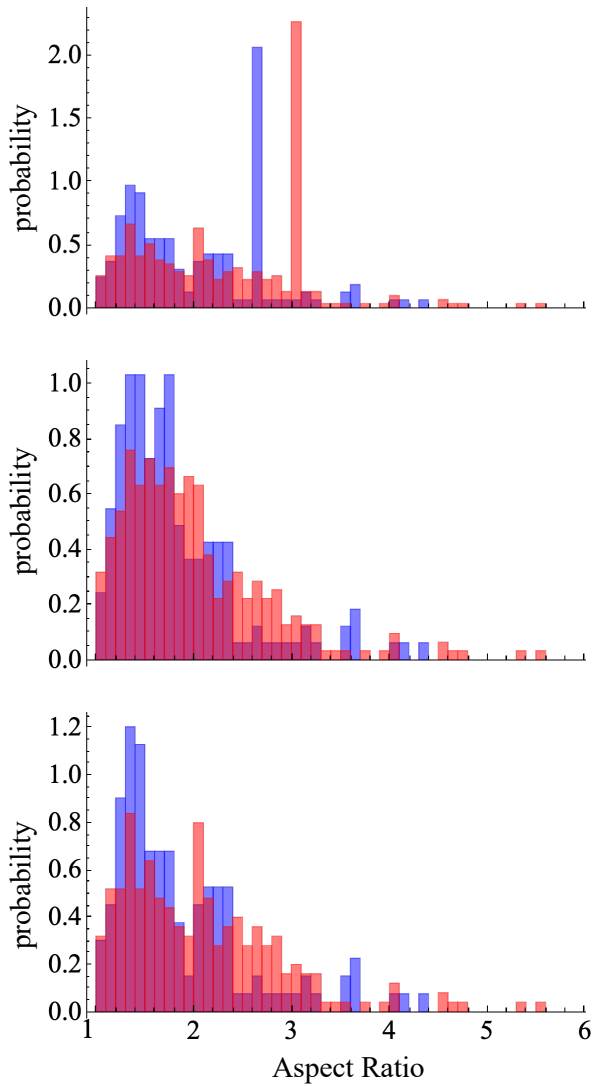

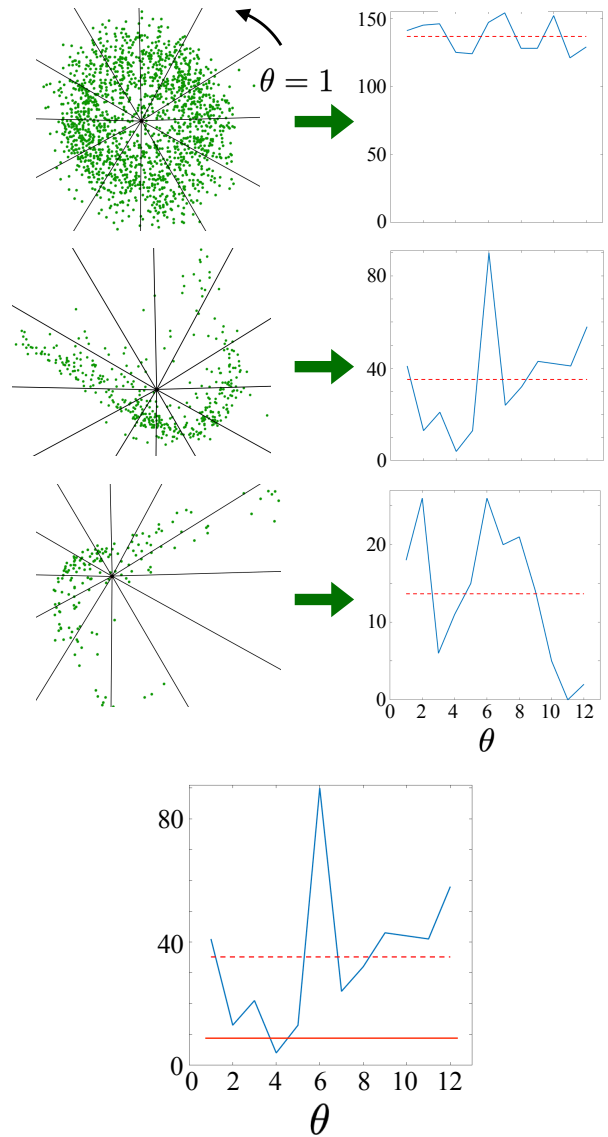

FIG. S6: Experimental aspect ratio cumulated probability distributions for all the conditions where proteins have low expression level and all the conditions where all proteins CD4, CCR5 and CXCR4 are over-expressed. Top: same as Figure 6 in the main text, except that the two peaks corresponding to the corrected AR of curled or coiled elongated nanodomains are represented here with their full heights. Middle: same distribution except that curled or coiled elongated nanodomains appear here with their raw AR, prior to correction. Bottom: same distribution except that curled or coiled elongated nanodomains do not appear at all. Only nanodomains not detected as curled or coiled have been pooled in these histograms. The $p$-values (KS statistical test, see below) are respectively $< 10^{-7}$, 0.00145 and 0.00143.

FIG. S7: Illustrative examples of the method used to identify the "false roundish" domains. Top: Analysis of the number of points per angular sector. The plane is divided into 12 regular angular sectors originating from the point-set barycenter, indexed by $\theta = 1$ to 12 (left); the number of points per sector is computed and plotted as a histogram (right). The average value of the number of points per sector is depicted with a red dashed line. Bottom: Domains for which at least one sector contains less than $t = 25\%$ of the average number of points per sector are considered to have a hole and then to be coiled elongated. The average value is still represented with a red dashed line, whereas the threshold $t$ is depicted with a red full line.

each experimental condition.

Since the sole AR is not sufficient for characterizing these domain shapes but we still wanted to take these "hidden" elongated domains into account, we assigned to them the average aspect ratio value of the initially detected elongated domains (verifying AR > 2). Figure 6 in the main text shows the modification of the AR distributions shown in Figure S6, after applying the correction described above. The correction is captured through two new peaks, around 2.6 in the distribution associated with a low level of expression, and 3.1 in the distribution associated with protein overexpression.

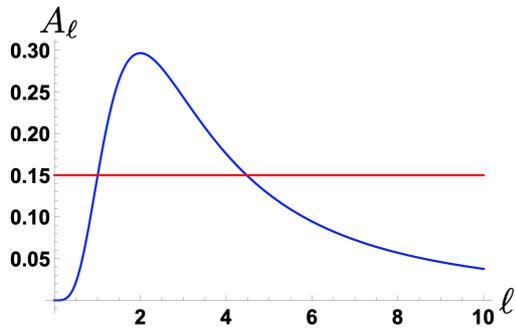

FIG. S8: Determination of the stability of elongated domains. In blue, plot of $A_\ell$ in fonction of the repulsion range $\ell$. The maximum coordinates are $(\ell^* = 2, A^* = \frac{8}{27})$. The red horizontal line represents $\frac{3}{2\pi\varepsilon}$. Ellipses are more stable than discs when the red line is below the blue curve.

We tested the two different threshold values, $t = 25\%$ and $30\%$, to detect the elongated domains, and checked that our conclusions were robust to modifications of the threshold value. We also performed the same analysis with an aspect ratio threshold $AR_0$ of 1.8 instead of 2 and also checked that the same conclusions hold with marginal influence on the trajectory classification.

## IV. STABILITY OF ELLIPSES WITH RESPECT TO DISCS

We discuss in the main text that ellipses are more stable than discs when

$$A_\ell > \frac{3}{2\pi\varepsilon} \tag{S5}$$

As illustrated in Figure S8, $A_\ell$ has a maximum $A^* = \frac{8}{27}$ at $\ell^* = 2$. Hence there exists an interval $[\ell_i^*, \ell_s^*]$ of values of $\ell$ where ellipses are stable if and only if $A^* > 3/(2\pi\varepsilon)$.

## V. SUPPLEMENTARY INFORMATION ON NUMERICAL SIMULATIONS

### A. Triangulated-vesicle model

We present briefly the numerical model developed and validated in Ref. [6]. It is a lattice-gas model, equivalent to the 2D Ising model with Ising interaction parameter $J_I$, describing the binary mixture of two species A and B. The species A can be considered as a forming a separated phase containing some membrane proteins and/or particular lipids. The lattice-gas model is coupled to a discretized Helfrich model [7] accounting for the membrane elasticity, with bending modulus $\kappa$ and surface tension $\sigma$. The membrane has the topology of a spherical vesicle, and is modeled as a triangulated surface. A composition variable, equal to A or B, is attached to each of the $N = 2562$ vertices of the triangulation. The Ising model is defined on this 2562-vertex lattice, which is a triangular lattice except for 12 vertices having 5 neighbors only. The local spontaneous curvature $C$ depends on the composition. The species A has a spontaneous curvature $C_A$ different from the majority phase (species B with $C_B = 2/R < C_A$ where $R$ is the average radius of the vesicle). We introduce the dimensionless parameters $c_1 = R(C_A - C_B) = RC_A - 2$, $\tilde\sigma = R^2\sigma/(k_BT)$ ($\sigma$ is the surface tension) and $\tilde J_I = J_I/(k_BT)$. Together with the reduced bending rigidity $\kappa/(k_BT)$ and the fraction $\bar\phi$ of species A vertices, they fully characterize the system.

The system undergoes two kinds of Monte Carlo (MC) moves: small radial displacements of the vertices and spin-exchange Kawasaki moves as far as composition is concerned. A total of more than $10^{10}$ MC steps is performed for each condition. The equilibration times have been measured in Ref. [6], and estimated to be shorter than $10^8$ MC moves for the regime of parameters studied in the present work. In general, simulations are run after starting from a perfect sphere on which species A and B are randomly distributed. To ascertain that the system under study are in equilibrium, we have run one simulation after starting from a single macro-cluster (Figure S9). We observe that the system has rapidly forgotten its initial state. The unstable macro-cluster splits into smaller fragments. Accordingly, the measured distributions of cluster sizes are essentially insensitive to the initial configuration. This demonstrates that small clusters are not just the fruit of slow coarsening when starting from a random configuration.

### B. Interacting-protein model

We also performed simulations on planar membranes with a different model of interacting "proteins" [8]. The mid-range repulsion between the proteins in this case is not explicitly promoted by the difference of spontaneous curvature as in the previous model, but comes from an effective pairwise potential, as if the membrane degrees of freedom were integrated out. Indeed, as explained in the main text, those two approaches are equivalent from a statistical physics perspective [9]. Both mechanisms make large domain unstable because the energy of a domain grows faster than its area, and promote mesophases. This also enables us to extend our analysis and results beyond the case of effective repulsion mediated by composition-shape coupling, e.g. interactions of electrostatic origin in soft matter physics [10, 11]. The pairwise potential reads [8, 10]:

$$U(r) = -\varepsilon_a e^{-\gamma_a r} + \varepsilon_r e^{-\gamma_r r} \tag{S6}$$

$r$ being the distance between the molecules. The parameters used for the simulations are $\varepsilon_a = 21.3 k_BT$, $\varepsilon_r = 0.31 k_BT$, $\gamma_a = 2$ and $\gamma_r = 0.25$ in inverse units of

6

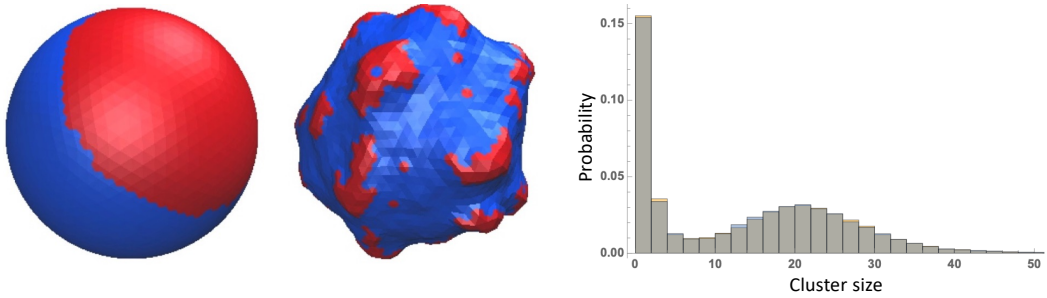

FIG. S9: If the simulations starts from a macro-cluster (left), one rapidly arrives (middle, example after $10^9$ MC moves) to the same kind of configuration as if one had started from a random distribution of species A and B. Right: Probability distributions of cluster sizes when starting from a random configuration (orange) or the macro-cluster (blue). They are almost super-imposed, indicating that the system is in equilibrium. Simulation parameters are the same as in Figure 5 in the main text, with $\bar{\phi} = 0.2$.

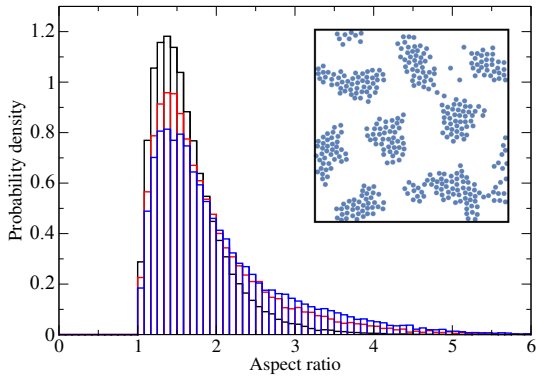

FIG. S10: Aspect ratio probability distributions for the domains observed in the in-plane simulations with explicit short-range attraction and long-range repulsion [8] as defined in Eq. (S6) with $\bar{\phi} = 0.18$ (black), $\bar{\phi} = 0.25$ (red) and $\bar{\phi} = 0.27$ (blue). The inset is a snapshot for $\bar{\phi} \simeq 0.25$ showing elongated domains (400 particles, periodic boundary conditions).

molecule diameter [1]. Figure S10 shows that this model also leads to the formation of more elongated domains when the concentration is high enough, whereas the domains are roundish at lower concentrations. Again the "hinge point" in the aspect ratio distributions is close to 2, which strengthens further our choice $AR_0 = 2$.

We now estimate the parameter $\varepsilon$ for this interacting-protein model. As compared to the main text, we have to estimate the parameters $\lambda$ and $E_0$ in a different way. We know the short-range attraction strength $\varepsilon_a$ and the mid-range repulsion strength $\varepsilon_r$ of the pairwise interaction, and we denote by $d$ their typical diameter (i.e. their typical separation). The line tension is then estimated to be $\lambda \approx \varepsilon_a/d$ and the repulsion parameter to be $E_0 \approx$ $\varepsilon_r/d^4$ [2]. In addition, we introduce the reduced diameter $\tilde{r}_0 = r_0/d$ (estimated to be close to 4 in the simulation outputs). It follows that $\lambda/(E_0 r_0^3) \approx \varepsilon_a/(\varepsilon_r \tilde{r}_0^3) \approx 1$ with the parameter values of our simulations. Again, we are close to the threshold, even though we do know the exact numerical prefactor in the stability condition for ellipses for the pairwise potential used in these simulation and the real density profile inside the clusters.

### C. Supplementary analyses

Regarding the triangulated-vesicle model, Figure S11 shows the cluster-size distribution for the two different concentrations $\bar{\phi}$ of the minority phase A studied in this work. Indeed, we implemented the cluster detection algorithm described in Ref. [6] to identify each domain and measure its size on independent system configurations throughout the simulation.

Figure S12 shows the spatial auto-correlation function, and the associated structure factor in the spherical harmonics basis, for same two concentrations in species A. In both cases, spatial wavelengths are very close. The oscillations of the spatial correlation functions are almost in phase and the two systems have a maximum in their structure factor for the same mode (order $l = 8$). The differences in amplitudes for the correlation functions and in the peak widths in the structure factors arise from the different input concentrations. Although the correlation function is computed from an average measurement in every direction, we can qualitatively interpret these observations as the fact that the domains elongate in one direction only and their typical spacing and thus width remain roughly constant. This has also been observed in the experimental data were the domain minor axis is weakly varying and their major axis is varying more significantly with concentration, as shown in Table S2. For

---

[1] The density $\bar{\phi}$ is computed at null temperature, with all the particles considered to be condensed in a perfect triangular lattice. They experience hard-core repulsion.

[2] Here we approximate the area occupied by each protein by $d^2$.

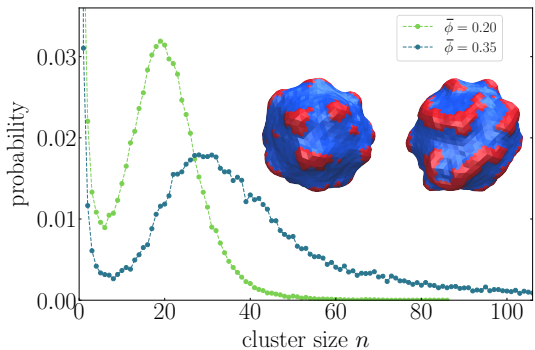

FIG. S11: Domain size probability distribution for vesicles with $\bar{\phi} = 0.20$ and $\bar{\phi} = 0.35$ in terms of number of sites. The typical domain size for a system corresponds to the position of the peak. The system at higher concentration ($\bar{\phi} = 0.35$) shows domains with a higher typical size (by $\approx 50\%$) than the one at lower concentration ($\bar{\phi} = 0.20$). Its largest domains are also bigger than the ones at lower concentration and can reach $\approx 400$ sites. Same parameters as in Figure 5 in the main text. Inset: simulations snapshots of the corresponding vesicles with $\bar{\phi} = 0.20$ (left) and $\bar{\phi} = 0.35$ (right); same snapshots as in Figure 5 in the main text.

example, in the condition where CCR5 is tracked and all protein expression levels are low, the minor axis of the domains was in average equal to $64 \pm 7$ nm, and it slightly shifted to $75 \pm 4$ nm in the case where all the proteins are overexpressed. In comparison, the major axis mean value shifted from $109 \pm 13$ nm to $144 \pm 7$ nm, a significantly more pronounced increase.

Note also that the analysis based upon the triangulated-vesicle model was additionally run for the same parameter values except a slightly different curvature coupling value as compared to the value $c_1 = 10$ examined in the main text. It led to very similar results: $c_1 = 9$ led to 24.5% of elongated domains at $\bar{\phi} = 0.2$ and 37.0% at $\bar{\phi} = 0.35$, whereas $c_1 = 11.0$ led to 31.1% of elongated domains at $\bar{\phi} = 0.2$ and 40.8% at $\bar{\phi} = 0.35$. This reinforces our conclusions on the effect of concentration on domain shape.

Even though coiled elongated structures as presented in section III above, that could not be detected by the sole aspect ratio criterion, were observed in the simulations for both numerical models (see Figure S13), this was limited to very high concentrations. They were exceptional in the concentrations used in Figures 5 in the main text and S10. For this reason, the classification correction of domain shapes based on angular distribution analysis was not applied there.

## VI. ABOUT $p$-VALUES

We have used several approaches to test whether two unpaired datasets arise from the same probability distribution through the $p$-value, as follows.

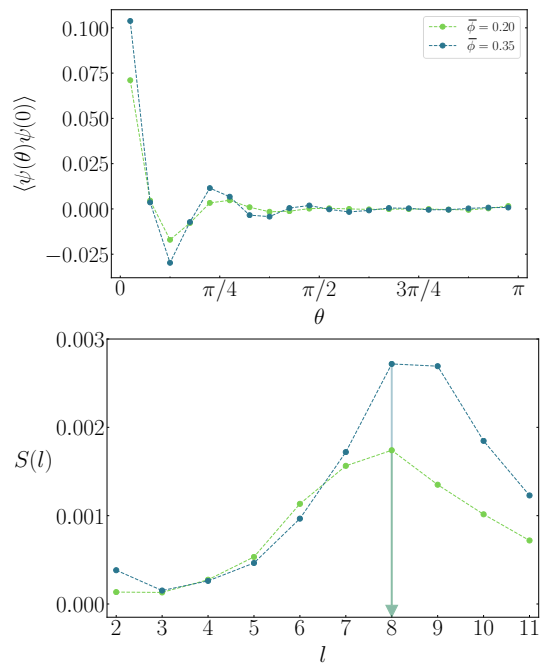

FIG. S12: Correlation functions (top) and associated structure factor (bottom) for the triangulated-vesicle numerical model, with $\bar{\phi} = 0.20$ and $\bar{\phi} = 0.35$. The parameter $l$ is the order of the spherical harmonics. The two systems are characterized by very close spatial wavelengths. Other simulation parameters are $c_1 = 8.0$, $\tilde{\sigma} = 300$ and $\tilde{J}_I = 0.5$.

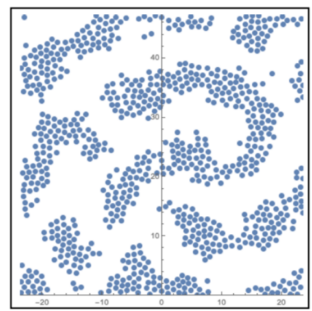

FIG. S13: Example of curved and curled domains obtained in the interaction-protein model for higher concentrations $\bar{\phi} = 0.31$ (800 particles, periodic boundary conditions).

Every transient confinement zone (TCZ) is considered a sample. Indeed, if several TCZs are extracted from the same cell, they do not in general correspond to the same nanodomain, and are thus uncorrelated in terms of cell surface topography.

We first computed the $p$-values based on the aspect ratio lists of two samples with the help of the Student t-test (TT) provided by Wolfram Mathematica (unpaired two-tailed t-test). However, the algorithm systematically displayed a warning because the distribution are not sufficiently close to Gaussian ones to ensure the validity of the Student test. We therefore switched to a more

| Test | $t$ | CD4 | CCR5 | CXCR4 |
|------|------|------|------|------|
| TT | 0.25 | 0.012 | $5 \times 10^{-5}$ | 0.005 |
| TT | 0.3 | 0.002 | $1 \times 10^{-4}$ | 0.001 |
| KS | 0.25 | $2 \times 10^{-4}$ | 0.002 | 0.001 |
| KS | 0.3 | $2 \times 10^{-6}$ | $8 \times 10^{-4}$ | $5 \times 10^{-4}$ |

TABLE S3: $p$-values obtained with different statistical tests and different values of the threshold $t$ in the case where the three proteins are overexpressed and one of them is tracked, as compared to the case where no protein is overexpressed and the same one is tracked, as in Figure 7 of the main text.

adapted non-parametric test, namely the Kolmogorov-Smirnov (KS) test also provided by the Wolfram Mathematica software. In addition, we tested two numerical values of the threshold $t$ used to detect false roundish domains. In the main text figures, a value of $t = 25\%$ is used. We have also calculated the $p$-values for an alternative value of 30%, as displayed in Table S3. A $p$-value $< 0.05$ is generally considered to be statistically significant [12]. Our KS $p$-values are on the order of, or lower than 0.001.

In all the figures, $*$ means $p \leq 0.05$, $**$ means $p \leq 0.01$, and $***$ means $p \leq 0.001$. When indicated on top of histograms displayed in insets, the $p$-values are calculated on the basis of the full distributions, not on the sole histograms that contain less information.

## VII. HIV-1 INFECTION ASSAYS

Infection assay were carried out as described in [2]. Briefly, the induced 293-Affinofile cells ($2 \times 10^5$ cells per well) in 96-well plates were inoculated with 100 ng p24 of the pNL4-3-derived viral clones expressing Renilla luciferase and gp160 from the JR-CSF or JR-FL strains. Infected cells were further incubated for 30 h at 37°C before being lysed. Viral replication was then determined by measuring luciferase activity in the cell lysates using a Glomax 96-well plate luminometer (Renilla Luciferase Assay, Promega, Madison, WI, USA) [13, 14].




\end{document}